\documentclass[a4paper,english, prb, final, floatfix, twocolumn]{revtex4}

\usepackage[ansinew]{inputenc}
\usepackage[english]{babel}
\usepackage{graphicx}
\usepackage{amsmath, amssymb, amsfonts,bm}
\usepackage{booktabs}
\usepackage{dcolumn}
\usepackage[colorlinks]{hyperref}
\usepackage[usenames]{color}
\usepackage[protrusion=true, expansion=true]{microtype}

\newcommand{\matr}[1]{{#1}}

\newcommand{\unitvekt}[1]{\hat{\bm{#1}}}
\newcommand{\erik}[1]{\mathrm{#1}}
\newcommand{\D}[1]{\,\text{d}#1}

\begin{document}
\preprint{BB-DomainWalls}

\title{Hard domain walls in superfluid ${}^3$He-B}

\author{Matti Silveri}
\altaffiliation{Present address: Department of Physics, Yale University, Connecticut 06520, USA}
\author{Tero Turunen}
\author{Erkki Thuneberg}
	\affiliation{Department of Physics, University of Oulu, P.O.Box 3000, FI-90014 Oulu, Finland}
\date{\today}

\begin{abstract}
We study theoretically planar interfaces between two domains of superfluid ${}^3$He-B. The structure of the B-B walls is determined on the scale of the superfluid condensation energy, and thus the domain walls have thickness on the order of the Ginzburg-Landau coherence length $\xi_{\rm GL}$. We study the stability and decay schemes of five inequivalent structures of such domain walls using one-dimensional Ginzburg-Landau simulation. We find that only one of the structures is  stable against small perturbations. We also argue that B-B interfaces could result from adiabatic A$\rightarrow$B transition and study textures at B-B interfaces. The B-B interface has a strong orienting effect on the spin-orbit rotation axis $\unitvekt{n}$ producing textures similar as caused by external walls.
We study the B-B interface in a parallel-plate geometry and find that the conservation of spin current sets an essential condition on the structure. The stable B-B interface gives rise to half-quantum circulation.
The energies of bound quasiparticle excitations are studied in a simple model.
\end{abstract}

\maketitle


\section{Introduction}
The internal interfaces of superfluid ${}^3$He can be divided into soft and hard ones.
Soft interfaces have  energy density on the order of the dipole-dipole interaction, and the interface thickness is on the order of  dipole length $\xi_{\rm D}\sim 10$ $\mu${\rm m}. Hard interfaces have higher energy density, which is on the order of the  superfluid condensation energy, and the interface thickness is much smaller, on the order of the Ginzburg-Landau~(GL) coherence length $\xi_{\rm GL}\sim 10$~nm. 
The soft interfaces are also known as solitons~\cite{Maki}. The solitons are mostly studied in the A phase (see Ref.\ \onlinecite{Hanninen03} for example). In B phase there is  evidence of the so-called $\theta$ solitons in the bulk\cite{smvp} and different types of solitons occur in restricted geometry \cite{Levitin13}. Of the hard interfaces, the best known case is the interface between the A and B phases \cite{abinterface, EVT2}. 
In the A phase hard interfaces have been theoretically discussed in restricted geometry \cite{Ohmi82}.  Hard interfaces in the B phase were discussed theoretically by Salomaa and Volovik~\cite{SV} who called them as ``cosmic\-like'' domain walls. There is renewed interest in these because two measurements that could be interpreted as evidence of B-B interfaces. In the first one, a phase shift of $\pi$ was observed for a superfluid loop including a weak link \cite{MAV}. In the second, anomalously high damping of a vibrating-wire resonator was observed \cite{WBG}. Narrow stripes of B-B interfaces are known to exist in the double-core vortex \cite{EVT} and similar structures appear in anisotextural Josephson effect \cite{VT02}.

In this paper we study several properties of hard B-B interfaces. These defects are non-topological and thus disappear by a sufficiently large perturbation. Using numerical solution of the Ginzburg-Landau equations we study the decay modes of the five structures suggested by Salomaa and Volovik~\cite{SV}. We find that only one of them is stable against small perturbations (Sec.\ \ref{s.stab}).  We discuss how a B-B interface could be generated by adiabatic cooling from the A phase  (Sec.\ \ref{s.adiab}). We study the texture around the B-B interface and find similar orienting effect  as near the walls of a superfluid ${}^3$He sample (Sec.\ \ref{s.dipole}). The contact line of the B-B interface with a wall is studied in  parallel-plate geometry, and an essential effect  of  spin currents on the structure is found (Sec.\ \ref{s.surf}). This problem is closely related to the striped phase found in Ref.\ \onlinecite{Vorontsov07}.
The stable B-B interface is found to generate superfluid circulation of a  half-quantum value (Sec.\ \ref{s.circ}).
The quasiparticle excitations bound to the interface are calculated using a simple model for the order parameter (Sec.\ \ref{s.excitation}).

\section{General properties of domain walls}
Let us consider generally an interface between two degenerate states. Often the intermediate states between these two have a higher energy. This energy cost leads to an interface of finite thickness, which we call a {\em domain wall}. The opposite case is that there is a continuum of degenerate states between the two states. In this case the interface tends to expand as thick as allowed by external conditions such as the homogeneity of the sample or external fields. In order to distinguish this case from the domain wall, we call it a {\em texture}.

The general order parameter of superfluid ${}^3$He is a $3\times3$ matrix $\matr{A}_{\mu i}$~\cite{Leggett}. We are interested in the energy scale of the superfluid condensation energy. On this scale, we can neglect the much smaller dipole-dipole energy. We also assume there is no external fields. In this case the order parameter in bulk B phase has the form~\cite{book}
\begin{equation}
\matr{A}_{\mu i}=\Delta \erik{e}^{\erik{i}\phi} \matr{R}_{\mu i}(\unitvekt{n},\theta).
\label{order}
\end{equation}
Here $\Delta$ is a scalar amplitude (real and positive), $\phi$ a phase (real) and $\matr{R}_{\mu i}$ a proper rotation matrix. The rotation matrix is real-valued, orthogonal ($\textsf{R}\textsf{R}^T=\textsf{1}$), and proper rotations also satisfy $\mathop{\rm det}\textsf{R}=1$. Proper rotations can be parametrized through rotation axis $\unitvekt{n}$ and angle $\theta$. Note that also improper rotations  ($\mathop{\rm det}\textsf{R}=-1$) could have been allowed, but it is more convenient to parameterize such a possibility by redefining the phase $\phi\rightarrow\phi+\pi$. The B phase is degenerate with respect to the variables $\phi$, $\unitvekt{n}$ and $\theta$. This degeneracy space is denoted by $\mathcal{R}=U(1)\times SO(3)$. Because this degeneracy space contains no disjoint pieces, there are no topological domain walls. Formally this is expressed using the homotopy group $\pi_0(\mathcal{R})=0$ (Ref.~\onlinecite{VM}). This means that hard B-B interfaces are nontopological, i.e. they can be broken by a sufficiently large perturbation.

In order to look for possible B-B interfaces, we  specify the order parameters on both sides of the interface. Let $\phi^{L}$ and $\matr{R}_{\mu i}^{L}$ denote the order parameter on the left hand side and
$\phi^{R}$ and $\matr{R}_{\mu i}^{R}$ on the right hand side. Let us consider some scalar observable describing the interface, for example the interface tension $\sigma$. Assuming it is a unique function of the two phases, it is a function of the $\phi^{L}$, $\matr{R}_{\mu i}^{L}$, 
$\phi^{R}$ and $\matr{R}_{\mu i}^{R}$. Similarly as in Ref.~\onlinecite{VT02}, symmetry allows to simplify this dependence: there is invariance to global phase shifts, to global spin rotations, and to rotations around the axis $x$ perpendicular to the interface. Therefore, $\sigma$ can only depend on three scalar invariants
\begin{equation}
\phi=\phi^{R}-\phi^{L},\ \  \psi_\perp=\matr{R}_{\mu x}^{L}\matr{R}_{\mu x}^{R},\ \ 
\psi_\parallel=\matr{R}_{\mu y}^{L}\matr{R}_{\mu y}^{R}+\matr{R}_{\mu z}^{L}\matr{R}_{\mu z}^{R},
\label{e.3inv}\end{equation}
where we have assumed summation over the repeated index $\mu$.
Alternatively, $\sigma$ can only depend on two numbers
\begin{equation}
\begin{split}
a&=e^{i\phi}\psi_\perp=e^{i(\phi^{R}-\phi^{L})}\matr{R}_{\mu x}^{L}\matr{R}_{\mu x}^{R},\\ 
b&=e^{i\phi}\psi_\parallel=e^{i(\phi^{R}-\phi^{L})}(\matr{R}_{\mu y}^{L}\matr{R}_{\mu y}^{R}+\matr{R}_{\mu z}^{L}\matr{R}_{\mu z}^{R}),
\end{split}
\label{e.2inv}
\end{equation}
which are complex valued in general. 

Only certain values of $a$ and $b$ (\ref{e.2inv}) can lead to stable interfaces. Values of $a$ and $b$ with a finite imaginary part would lead to mass current through the interface. Such structures could be stabilized by an external mass-current bias, but otherwise they would relax to currentless states.  Similar conclusion applies to most real values of $a$ and $b$: these give rise to spin current through the interface and the resulting structure would be unstable in the absence of external spin-current bias. The spin current through the interface can vanish only for certain symmetric cases corresponding to $a=\pm 1$, and $b=0$, $\pm2$. The resulting six cases are listed  in Table \ref{TauluKoonti}.
The interfaces are  labeled by $ab$, where a minus sign of $a$ or $b$ is  indicated by a bar over the number. For example, $\overline{1}2$ implies $a=-1$ and $b=2$.  
\begin{table*}
\begin{center}
\begin{scriptsize}
\begin{tabular}{cccccclccccc}
\toprule
Name&$n$ &$\phi$&$\psi_\perp$&$\psi_\parallel$&  $\widetilde{\textsf{A}}^R$ & $H_P$&Type& $H_S$& $\sigma_{\text{calc}}/{f_{\rm c}^{\rm B}\xi_{\rm GL}}$&$\sigma_{\text{teor}}/{f_{\rm c}^{\rm B}\xi_{\rm GL}}$&Fig.\\\hline\\[-1mm]
$12$&$0$&$0$&$1$&$2$& $\begin{bmatrix} +1&0&0\\0&+1&0\\0&0&+1 \end{bmatrix}$ &$\infty ,m, T  $ &\begin{large}$\substack{\text{Bulk}\\\text{(no interface) }}$ \end{large}& \begin{large}$\substack{\infty ,m ,T}$ \end{large}&\begin{footnotesize}$0$\end{footnotesize}&\begin{footnotesize}$0$\end{footnotesize}\\\\
$10$&$2,3$&$\pi$&$-1$&$0$& $\begin{bmatrix} +1&0&0\\0&{-}1&0\\0&0&+1 \end{bmatrix}$&\begin{large}$\substack{2_x,m_y,m_z,T,2_y^o}$\end{large} &\begin{large} $\substack{\text{Single}}$ \end{large}& $2_x,m_y,m_z, T, 2_y^o$&\begin{footnotesize}$0.9000$\end{footnotesize}&\begin{footnotesize}$0.9000$\end{footnotesize}&\ref{BB10.fig}(a)\\\\
$\overline{1}2$&$1$&$\pi$&$1$&$-2$& $\begin{bmatrix} {-}1&0&0\\0&+1&0\\0&0&+1 \end{bmatrix}$ &$\infty_x,m_y, m_z, T, m_x^o $ &\begin{large}$\substack{\text{Mixed}}$\end{large}& \begin{large}$\substack{T}$\end{large} &\begin{footnotesize}$1.0797 $\end{footnotesize}&\begin{footnotesize}$0.9000$\end{footnotesize}
&\ref{BBm12.fig}(b)\\\\
$1\overline{2}$&$6$&$0$&$1$&$-2$&$\begin{bmatrix} +1&0&0\\0&{-}1&0\\0&0&{-}1 \end{bmatrix}$& $\infty_x, m_y, m_z, T ,m_x^o $ &\begin{large}$\substack{\text{Double} \\\\ \text{Texture}}$\end{large}& \begin{large}$\substack{2_x,m_y,m_z,T \\\\ \infty_x, T}$ \end{large} &\begin{large}  $\substack{1.7998 \\\\0.0953}$ \end{large} &\begin{large}$\substack{1.8000 \\\\0}$\end{large}&\begin{large}$\substack{\ref{BB1m2.fig}(b)\\\\ \ref{BB1m2.fig}(c)} $\end{large} \\\\
$\overline{1}0$&$4,5$&$0$&$-1$&$0$&  $\begin{bmatrix} {-1}&0&0\\0&{-1}&0\\0&0&+1 \end{bmatrix}$& $ 2_x, m_y, m_z, T, 2_z^o$ &\begin{large}$\substack{\text{Mixed double} \\\\ \text{Mixed} \\\\ \text{Texture}}$\end{large}& \begin{large} $\substack{ m_y,T \\\\ 2_x,m_y,m_z \\\\ m_z, T}$ \end{large}& \begin{large} $\substack{1.9809  \\\\1.1333 \\\\0.1878}$ \end{large}&\begin{large}$\substack{1.8000 \\\\0.9000 \\\\0}$\end{large}&\begin{large}$\substack{\ref{BBm10.fig}(e)\\\\ \ref{BBm10.fig}(d)\\\\ \ref{BBm10.fig}(b)} $\end{large}\\\\
$\overline{1}\overline{2}$&$7$&$\pi$&$1$&$2$&$\begin{bmatrix} {-}1&0&0\\0&{-}1&0\\0&0&{-}1 \end{bmatrix}$ & $\infty_x ,m_y ,m_z ,T ,m_x $&\begin{large}$\substack{\text{Mixed triple} \\\\ \text{Mixed 2} \\\\ \text{Texture}}$\end{large}& \begin{large}$\substack{ T \\\\ T \\\\ \infty_x, m_y,m_z }$ \end{large} & \begin{large}  $\substack{2.8732  \\\\1.0890  \\\\0.2349 }$ \end{large} &\begin{large}$\substack{2.7000 \\\\0.9000 \\\\0}$\end{large}&\begin{large}$\substack{\ref{BBm1m2.fig}(f)\\\\ \ref{BBm1m2.fig}(g)\\\\ \ref{BBm1m2.fig}(c)} $\end{large}\\\\ 
\bottomrule
\end{tabular}
\end{scriptsize}
\caption{\label{TauluKoonti} Summary of different interface structures. The first column gives the name of the interface based on the invariants \eqref{e.2inv}. The second column gives the number of the interface according to Ref.~\onlinecite{SV}. The following three columns give the invariants $\phi$, $\psi_\perp$ and $\psi_\parallel$ (\ref{e.3inv}). The sixth column gives a possible choice for the reduced order parameter $\widetilde{\matr{A}}$ on the right hand side of the interface \eqref{e.atildr}. On the left hand side $\widetilde{\matr{A}}$ equals to the unit matrix in all cases. The $H_P$ column gives the elementary symmetry operators that generate the symmetry group of the domain wall problem. The elementary symmetry operators of converged interface configurations are given in column $H_S$. For notation see the main text. The column Type describes the different converged solutions. Column $\sigma_{\text{calc}}$ gives the calculated interface free energy for an arbitrarily fixed $L=140\; \xi_{\rm GL}$ calculation interval and $\sigma_{\text{teor}}$ refers to the theoretical value as $ L \rightarrow \infty$.  The last column gives the figure showing the converged order parameter. The first row refers to the bulk B phase, in the absence of any interface.}
\end{center}
\end{table*}

For comparison, Salomaa and Volovik~\cite{SV} studied numerically 7 types of B-B interfaces. Only five of them are different so that they have different invariants $\phi$, $\psi_\perp$ and $\psi_\parallel$ \eqref{e.3inv}. The two extra ones are duplicates that can be obtained by a rotation around the interface normal. 
The numbering of the structures according to Ref.\ \onlinecite{SV} is given in the second column of Table \ref{TauluKoonti}.

In order to represent the order parameter components, we define a reduced order parameter 
$\widetilde{\matr{A}}_{ij}$ by the relation
\begin{equation}
\matr{A}_{\mu j}(\bm r)=\matr{A}_{\mu i}^{L} \widetilde{\matr{A}}_{ij}(\bm r).
\label{reduc1}
\end{equation}
This implies that on the left hand side $\widetilde{\matr{A}}_{ij}(\bm r)$ reduces to a unit matrix, 
$\widetilde{\matr{A}}_{ij}^L=\delta_{ij}$. On the right hand side $\widetilde{\matr{A}}_{ij}(\bm r)$ reduces to the matrix \begin{equation}
\widetilde{\matr{A}}_{ij}^R=\frac1{\Delta^2}\matr{A}_{\mu i}^{L*}\matr{A}_{\mu j}^{R}
=\erik{e}^{i(\phi^R-\phi^L)}\matr{R}_{\mu i}^{L}\matr{R}_{\mu j}^{R},
\label{e.atildr}\end{equation}
which is closely related to the invariants $\phi$, $\psi_\perp$ and $\psi_\parallel$ (\ref{e.3inv}).

In the following we assume that the interface is homogeneous in its plane. This means that the order parameter $\widetilde{\matr{A}}_{\mu i}(x)$ only depends on the coordinate $x$ perpendicular to the the interface.
The interface problem has also additional symmetries, which may be helpful to understand the results obtained below. The different symmetries of both the problem (the energy functional and boundary conditions) and the solutions (the order parameter) are given in Table~\ref{TauluKoonti} using the following notation. Twofold and continuous rotation symmetry are denoted by 2 and $\infty$, respectively, and reflection is denoted by $m$. The subscripts indicate the axis to which each symmetry refers.  In the bulk (first row of Table~\ref{TauluKoonti}) the rotation and reflection symmetries are valid along any axis. Time inversion is denoted by $T$.  Generally the reflections and rotations refer to simultaneous operations in both spin and orbit spaces.  However, there are also symmetry operations that refer to orbit space only, and these are denoted with superscript $o$. For example, $m_x^o$ means the symmetry operation
$\matr{A}_{\mu i}(x)=e^{ic_2}\matr{A}_{\mu j}(-x+c_1)\matr{S}^x_{ji}$, where $\matr{S}^x_{ji}$ is a diagonal matrix with elements -1, 1 and 1 and $c_1$ and $c_2$ are some constants depending on the choice of coordinate axes and phases. The column $H_P$ of Table \ref{TauluKoonti} lists some elementary symmetry operations of the interface problem in each case. The complete symmetry group of the {\em problem} consists of all combinations of the elementary symmetry operations. The interface solutions ${\matr{A}}_{\mu i}(x)$ have either the same symmetry, or lower symmetry, the latter case being known as broken symmetry. The elementary symmetry operations, out of which the whole symmetry group of each {\em solution} can be constructed, are given in column $H_S$.

\section{Ginzburg-Landau calculation}\label{s.GL}
The interface structures are calculated using Ginzburg-Landau~(GL)~theory. This is valid in the  temperature region $T_c-T\ll T_c$ near the critical temperature $T_c$ of superfluid ${}^3$He. The free energy functional~$F=F_{\rm b}+F_{\rm g}$ consists of the bulk
\begin{equation}
\begin{split}
F_{\rm b}=\int\D^3r \big \{&f_{\rm c}^{\rm B}-\alpha \text{Tr}(\textsf{A}\textsf{A}^\dagger)+\beta_1|\text{Tr}(\textsf{A}\textsf{A}^{\matr{T}} )|^2\\&+\beta_2[\text{Tr}(\textsf{A}\textsf{A}^\dagger )]^2+\beta_3\text{Tr}(\textsf{A}\textsf{A}^{\matr{T}}\textsf{A}^* \textsf{A}^\dagger )\\ &+\beta_4\text{Tr}(\textsf{A}\textsf{A}^\matr{\dagger} \textsf{A}\textsf{A}^\dagger )+\beta_5\text{Tr}(\textsf{A}\textsf{A}^\dagger \textsf{A}^*\textsf{A}^{\matr{T}} )\big\}
\end{split}
\label{functionalbulk}
\end{equation}
and gradient energies 
\begin{equation}
\begin{split}
F_{\rm g}=K\int\D^3r \left[(\gamma-1)\partial_i \matr{A}_{\mu i}^*\partial_j \matr{A}_{\mu j}
+\partial_i \matr{A}_{\mu j}^*\partial_i \matr{A}_{\mu j}\right].
\label{functional}
\end{split}
\end{equation}
The GL theory has the input parameters $\alpha, \beta_i,K \text{ and } \gamma$. Our numerical results depend on only the value of $\gamma$ and the relative values of $\beta_i$. For the stability studies we mainly use the weak coupling values ($\gamma=3$, $-2\beta_1=\beta_2=\beta_3=\beta_4=-\beta_5$), but we also use the Sauls-Serene $\beta_i$'s for some tests~\cite{SS}. These tests support the conclusion that our stability results remain approximately unchanged within the stability region of the B phase. When quoting pressure, it is according to the theoretical scale of Ref.\ \onlinecite{SS}, where the polycritical pressure (2.85 MPa) is somewhat larger than the measured one. We take zero pressure to correspond to weak coupling values of $\beta_i$'s, and use linear interpolation between 0 and 1.2 MPa.

The GL theory fixes the amplitude of the order parameter in Eq.\ \eqref{order} to the value 
\begin{equation}
\Delta^2=\frac{\alpha}{6\beta_{12}+2\beta_{345}},
\label{delta}
\end{equation}
where $\beta_{ij}=\beta_i+\beta_j$ etc. The GL coherence length is defined by $\xi^2_{\rm GL}=K/\alpha$. The superfluid condensation energy density of the B phase is $f_{\rm c}^{\rm B}=\frac{3}{2}\alpha \Delta^2$.  This energy is added in $F_{\rm b}$ \eqref{functionalbulk} implying that the energy of the bulk B phase is adjusted to the zero. With this operation the energy~$F$ measures the deviation from the bulk phase. As the interface is assumed homogeneous, the interface free energy equals the tension
\begin{equation}
\sigma=\frac{F}{A},
\label{sigma}
\end{equation}
where $A$ is the area of the interface. Thus only one-dimensional integration is needed in equations \eqref{functionalbulk} and  \eqref{functional}. It is convenient to express $\sigma$ in units of  ${f_{\rm c}^{\rm B}\xi_{\rm GL}}$. Note that this unit is by factor 3/2 larger than the  one used in Ref.~\onlinecite{SV}. 

The minimization $F$ in one dimension leads to the differential equations
\begin{widetext}
\begin{equation}
\begin{split}
0=&\partial_x \partial_x [\matr{A}_{\mu i}+(\gamma-1)\matr{A}_{\mu i}\delta_{ix}]\\&-\left(-\alpha \matr{A}_{\mu i}+2\beta_1 \matr{A}^*_{\mu i} \matr{A}^*_{\nu j}\matr{A}_{\nu j}+2\beta_2 \matr{A}_{\mu i} \matr{A}^*_{\nu j}\matr{A}_{\nu j}+2\beta_3 \matr{A}^*_{\nu i} \matr{A}_{\mu j}\matr{A}_{\nu j}+2\beta_4 \matr{A}_{\nu i} \matr{A}^*_{\nu j}\matr{A}_{\mu j}+2\beta_5 \matr{A}_ {\nu i} \matr{A}_{\nu j}\matr{A}^*_{\mu j} \right)
\end{split}
\label{diffeq}
\end{equation}
\end{widetext}
for all indices $\mu$~and~$i$. At the end points $x=0$ and $x=L$ we fixed the values of $\matr{A}_{\mu i}$. Alternatively, a zero-derivative boundary condition could be applied. If the resulting interface structure is a single domain wall, the solution is independent of the length $L$ and of the type of the boundary condition as long as $L\gtrsim 20\;  \xi_{\rm GL}$.  If it splits into two or more domain walls, a longer $L$ is needed. However, if the result is a texture, it will depend essentially on the boundary condition even in the limit $L\rightarrow\infty$. Below we give the interface structures and energies for  $L=140\; \xi_{\rm GL}$ with fixed values of $\matr{A}_{\mu i}$ at the both ends. Discretization length $\Delta x= 1\; \xi_{\rm GL}$ was used in the presented stability results but the energies in Table \ref{TauluKoonti} were calculated by using $\Delta x= 0.25\; \xi_{\rm GL}$.

\begin{figure*}[ht!]
\centering 
\includegraphics[width=0.80\linewidth]{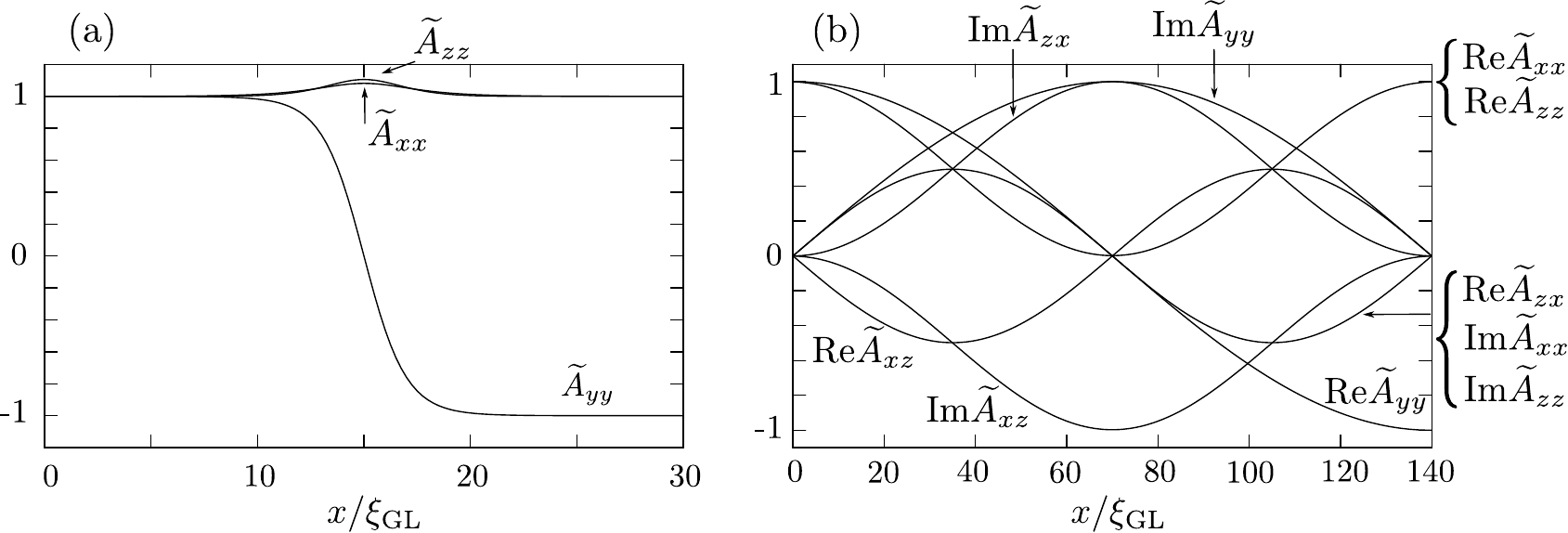}
\caption{\label{BB10.fig} Two solutions for the order parameter in the B-B-$10$ interface. The domain wall (a) is locally stable and decays to a texture (b) only as a result of a large perturbation. Note that the texture solution is independent on where the initial interface is located in the interval $0<x < 140 \xi_{\rm GL}$.  In this and the following figures, all non-vanishing real and imaginary parts of $\widetilde{\matr{A}}_{ij}$  are plotted.}
\end{figure*}
As initial values we used simple forms of $\widetilde{\matr{A}}_{ij}(x)$ consistent with the problem symmetries. The boundary value problem was solved iteratively using a simple relaxation method~\cite{EVT}. The iteration proceeds towards lower energy assuming the iteration step is kept small. The development under iteration can be considered qualitatively equal to time evolution. The iteration was continued as long as a converged solution was found. The stability of the converged solutions was tested by adding a perturbation $\delta\widetilde{\matr{A}}_{ij}(x)$ to $\widetilde{\matr{A}}_{ij}(x)$ and continuing the iteration. The perturbations had the form
\begin{equation}
\delta\widetilde{\matr{A}}_{ij}(x)=\sum_{k=0}^N\frac{\matr{C}^{(k)}_{ij}}{\cosh[(x-x_k)/s]}
\label{e.perturb}
\end{equation}
with parameters $\matr{C}^{(k)}_{ij}$, $x_k$, $N$ and $s$. The strength  of a perturbation $C$ is defined as the largest deviation from the unperturbed configuration, 
\begin{equation}
C=\max_{\stackrel{x \in [0,L]}{i,j}} \delta\widetilde{\matr{A}}_{ij}(x).
\label{C}
\end{equation}
The parameter $s$ measuring the width of the perturbation has typically value of $5\ \xi_{\rm GL}$, the shifting constants $x_k$ and the number $N$ of simultaneous perturbation peaks are case dependent. Typically 10 to 30 different perturbations were tested.

In the case the solution was found stable against small perturbations, we tested larger perturbations. An activation barrier can be determined by looking for the smallest energy perturbation that is sufficient to initiate convergence to a different type of solution. Note however, that this activation energy is one-dimensional (unit ${\rm J}/{\rm m^2}$), whereas the true decay takes place in the three dimensions.

\section{Stability and decay schemes}\label{s.stab}
The following five subsections give the stability and decay schemes of the domain walls, starting from the simplest case and proceeding to more complicated cases. The results are summarized in Table~\ref{TauluKoonti}. 
   
\subsection*{B-B-$10$}\label{s.BB10}
An example order parameter for a converged solution of interface 10 is shown in Fig.~\ref{BB10.fig}(a). This solution shows a well defined domain wall, and is similar as found earlier~\cite{SV}. It is characterized by a sign change in the component $\widetilde{\matr{A}}_{yy}$. The only other nonzero components are $\widetilde{\matr{A}}_{xx}$ and $\widetilde{\matr{A}}_{zz}$. They are almost constants, but are slightly enhanced at the domain wall. It can be noticed that there is a small difference between $\widetilde{\matr{A}}_{xx}$ and $\widetilde{\matr{A}}_{zz}$ at the domain wall. This is a consequence of the gradient energy~\eqref{functional}, which is $\gamma \approx 3$ times  more costly for $\widetilde{\matr{A}}_{i x}$, ($i=x,y,z$), than for the other components $\widetilde{\matr{A}}_{i y}$ and $\widetilde{\matr{A}}_{i z}$. 

The vanishing of $\widetilde{\matr{A}}_{\mu y}$ at the domain wall implies a {\em gap node} in the $y$ direction.
The quasiparticle excitations are studied in more detail in Sec.\ \ref{s.excitation}. Note that the gap node is in the plane of the domain wall.  
Thus a continuous set of degenerate domain walls can be formed by rotation around the interface normal. 
These include the interfaces 2 and 3 of Ref.~\onlinecite{SV}. The interface free energy $\sigma$  is $0.90\; f_{\rm c}^{\rm B} \xi_{\rm  GL}$ in the weak coupling but decreases to $0.83\;f_{\rm c}^{\rm B} \xi_{\rm  GL}$ at the pressure of $28$ bar. The presence of the domain wall changes the magnetic-susceptibility tensor of the B phase by an additional contribution $\matr{\chi}_{\mu \nu}=g_z\Delta^2\xi_{\rm GL}{\matr{R}}_{\mu i}^L\widetilde{\matr{\chi}}_{ij}{\matr{R}}_{\nu j}^L$. For B-B-$10$ the reduced susceptibility is  diagonal and has components
\begin{eqnarray}
\widetilde\chi_{xx}=-1.03, & \widetilde\chi_{yy}=+4.08, & \widetilde\chi_{zz}=-1.02.
\label{BB10susc}
\end{eqnarray}
The interface has spin current $J_{\mu i} =R_{\mu \nu}\widetilde J_{\nu i}$, where
\begin{align}
\widetilde J_{z y}=&\frac{2K\Delta^2}{\hbar}\int dx\left(\widetilde A_{yy}\partial_x \widetilde A_{xx}-\widetilde A_{xx}\partial_x \widetilde A_{yy}\right)\notag \\
=&1.127\times{4K\Delta^2}/{\hbar}
\label{e.spiczy}\end{align}
and all other components of $\widetilde J_{\nu i}$ vanish. Independently of the chosen coordinates, the spin current tensor can be written as $\textsf J=c\textsf{R}\hat{\bm u}\,\hat{\bm t}$, where $\hat{\bm u}=\hat{\bm s}\times\hat{\bm t}$, $\hat{\bm s}$ is the interface normal pointing towards the B phase (\ref{order}) of proper rotation matrix $\textsf{R}$, $\hat{\bm t}$ is the direction of the interface gap node, and  $c=-1.127\times4K\Delta^2/\hbar$. The  spin current arises from the filled states below the Fermi energy as pointed out  by Salomaa and Volovik \cite{SV} and is further discussed in Sec.\ \ref{s.excitation}. 

The domain wall structure has been calculated also by Vorontsov and Sauls using the weak coupling quasiclassical theory~\cite{VorSau}. This theory is not limited in temperature, and thus allows to generalize the present GL calculations to the whole temperature range $0\le T<T_c$.

Our numerical calculations indicate that B-B-10 is a local minimum of the free energy, i.e.\ after small perturbations the same converged solution was reached. Only a strong perturbation of amplitude~$C \gtrsim 0.7$ leads to a different solution. We looked for the smallest energy perturbation that leads the iteration away from the original solution.
The critical perturbation  energy  is roughly $8.7\;f_{\rm c}^{\rm B} \xi_{\rm  GL}$ and corresponds to  $s=3.5\; \xi_{\rm  GL}$ \eqref{e.perturb} in the weak coupling. The critical perturbation energy increases with increasing pressure and is $10.7f_{\rm c}^{\rm B}\; \xi_{\rm  GL}$ at $28$ bar. The optimal amplitude $C$ and the width of the perturbation $s$ stay unchanged. As the interface energy simultaneously decreases, this indicates higher stability of the  B-B-10 interface at high pressures, at least within the one dimensional approximation. 
\begin{figure*}
\centering
\includegraphics[width=0.75\linewidth]{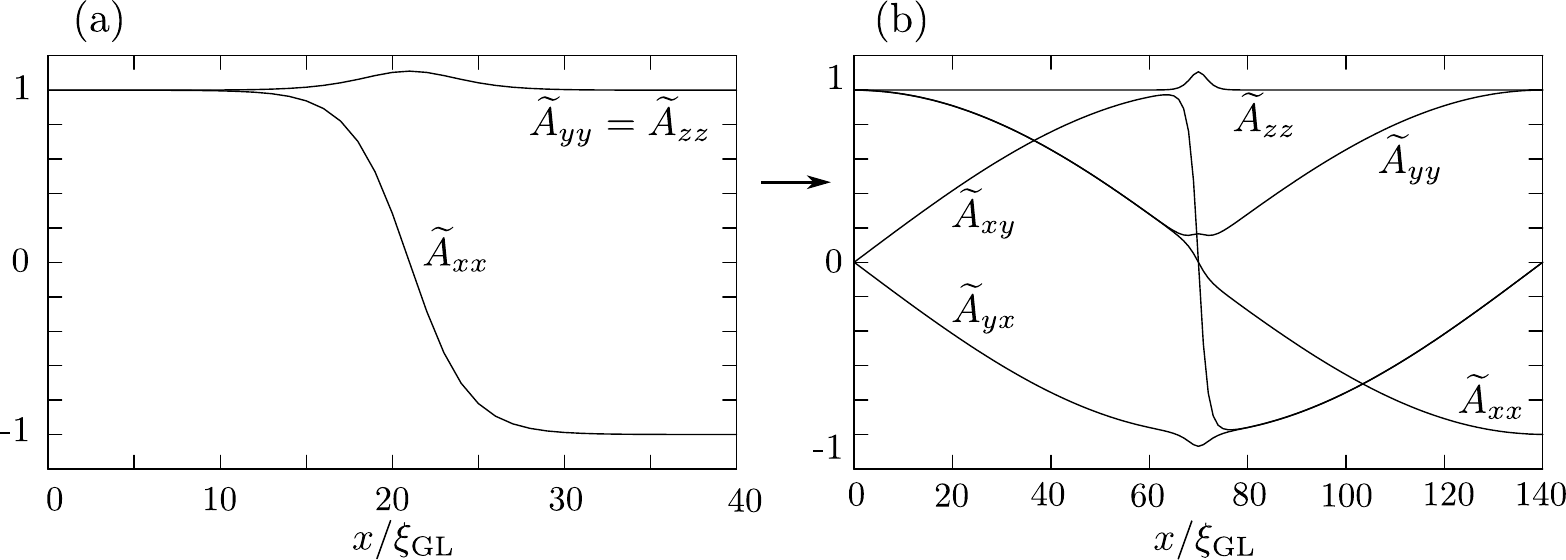} 
\caption{\label{BBm12.fig} Two solutions for B-B-$\overline{1}2$. The single domain wall (a) corresponds to a saddle point of energy and decays to a mixed domain wall (b).  Note the different  scales of $x$ in the panels. The domain wall (a) is wider than the one in Fig.\ \ref{BB10.fig}(a). In the middle of the mixed solution (b) one recognizes the B-B-$10$ domain wall in the abrupt change of $\widetilde{\matr{A}}_{xy}$.}
\end{figure*}

One possible converged solution that results from a large perturbation is shown in Fig.~\ref{BB10.fig}(b).
It also has an analytic description 
\begin{equation}
\widetilde{\matr{A}}_{ij}(x)=\erik{e}^{\erik{i} \pi x/L}\matr{R}_{ij}(\unitvekt{y},-\pi x/L),
\label{bb2text}
\end{equation}
where, as above, $\matr{R}_{ij}$ is a rotation matrix parameterized by an axis and a rotation angle.
This is a texture that has winding both in phase and in spin-orbit rotation.  Its  symmetry differs from the initial interface structure, in particular, the time reversal symmetry is broken. As could be guessed, this state is most easily generated by imaginary perturbations in the amplitudes $\widetilde{\matr{A}}_{xz}$, $\widetilde{\matr{A}}_{zx}$ and $\widetilde{\matr{A}}_{yy}$, centered at the domain wall. As this solution has broken symmetry, there exists other texture states with the same energy, which can be reached by a different type of perturbation. In contrast to the domain wall, the texture solution \eqref{bb2text} expands as long as allowed by the size $L$ of the calculation region. The texture solution is stabilized by requiring fixed values of $\widetilde{\matr{A}}_{ij}$ at the ends. If zero-derivative boundary conditions would have been used, all variation would move out of the calculation interval during the iteration, and the converged result  would be a constant order parameter.

\subsection*{B-B-$\overline{1}2$}
The first converged solution of the $\overline{1}2$ interface is shown in Fig.~\ref{BBm12.fig}(a). The domain wall is characterized by a sign change in the component $\widetilde{\matr{A}}_{xx}$ while
$\widetilde{\matr{A}}_{yy}$ and $\widetilde{\matr{A}}_{zz}$ are equal and nearly constants. The domain wall is thicker  than the $10$ domain wall [in Fig.~\ref{BB10.fig}(a)], which can be understood by the anisotropy of the gradient energy \eqref{functional} discussed above. The same effect  at a general temperature is discussed in Ref.~\onlinecite{VorSau}. The interface free energy $\sigma$  is $1.5403\;f_{\rm c}^{\rm B} \xi_{\rm  GL}$, which is higher than for the B-B-$10$ domain wall.

This domain wall is closely related to the structure at a planar wall. Namely if one requires vanishing of $\widetilde{\matr{A}}_{xx}$ and zero normal derivative of $\widetilde{\matr{A}}_{yy}$ and $\widetilde{\matr{A}}_{zz}$ at the wall, the order parameter is the same as in Fig.~\ref{BBm12.fig}(a) on one side of the interface. This boundary condition indeed is valid if the quasiparticles hitting the wall are reflected specularly~\cite{AGR}. This ideal case has been calculated repeatedly since the early work of Buchholtz and Zwicknagl~\cite{BJ}.

In contrast to the domain wall B-B-$10$, we find that the domain wall $\overline{1}2$ is a saddle point of energy.
A perturbation as small as $C= 10^{-40}$ is sufficient to cause convergence to another solution. The essential thing is that there is a symmetry in the original solution, which has to be broken in order to achieve lower energy.  One converged solution is shown in Fig.~\ref{BBm12.fig}(b) and it can be represented as 
\begin{equation}
\widetilde{\matr{A}}_{i j}(x)=\matr{R}_{ik}(\unitvekt{z},-\pi x/L) \widetilde{\matr{A}}_{kj}^{\text{\ref{BB10.fig}(a)}}(x-c),
\label{BB1}
\end{equation}
where matrix $\widetilde{\matr{A}}_{kj}^{\text{\ref{BB10.fig}(a)}}(x)$ is the B-B-$10$ order parameter shown
in Fig.~\ref{BB10.fig}(a) and $c$ is a constant. This is a mixed solution of the texture characterized by smooth rotation around $z$-axis and the B-B-$10$ domain wall in the middle. This corresponds to broken symmetry, in particular the continuous rotational symmetry of the domain wall is broken. There is large degeneracy of this state, and the particular state selected depends on the perturbation given to the domain wall solution. 

After a strong perturbation $C \gtrsim 0.6$, the single interface can deform to pure texture similar to the texture of the B-B-$10$ [Eq.\ \eqref{bb2text} and Fig.~\ref{BB10.fig}(b)], only the rotation axis is changed from $\unitvekt{y}$ to $\unitvekt{x}$. The perturbations leading to this solution can be designed analogously to the case of B-B-$10$.

\subsection*{B-B-$1\overline{2}$}

\begin{figure*}
\centering
\includegraphics[width=0.75\linewidth]{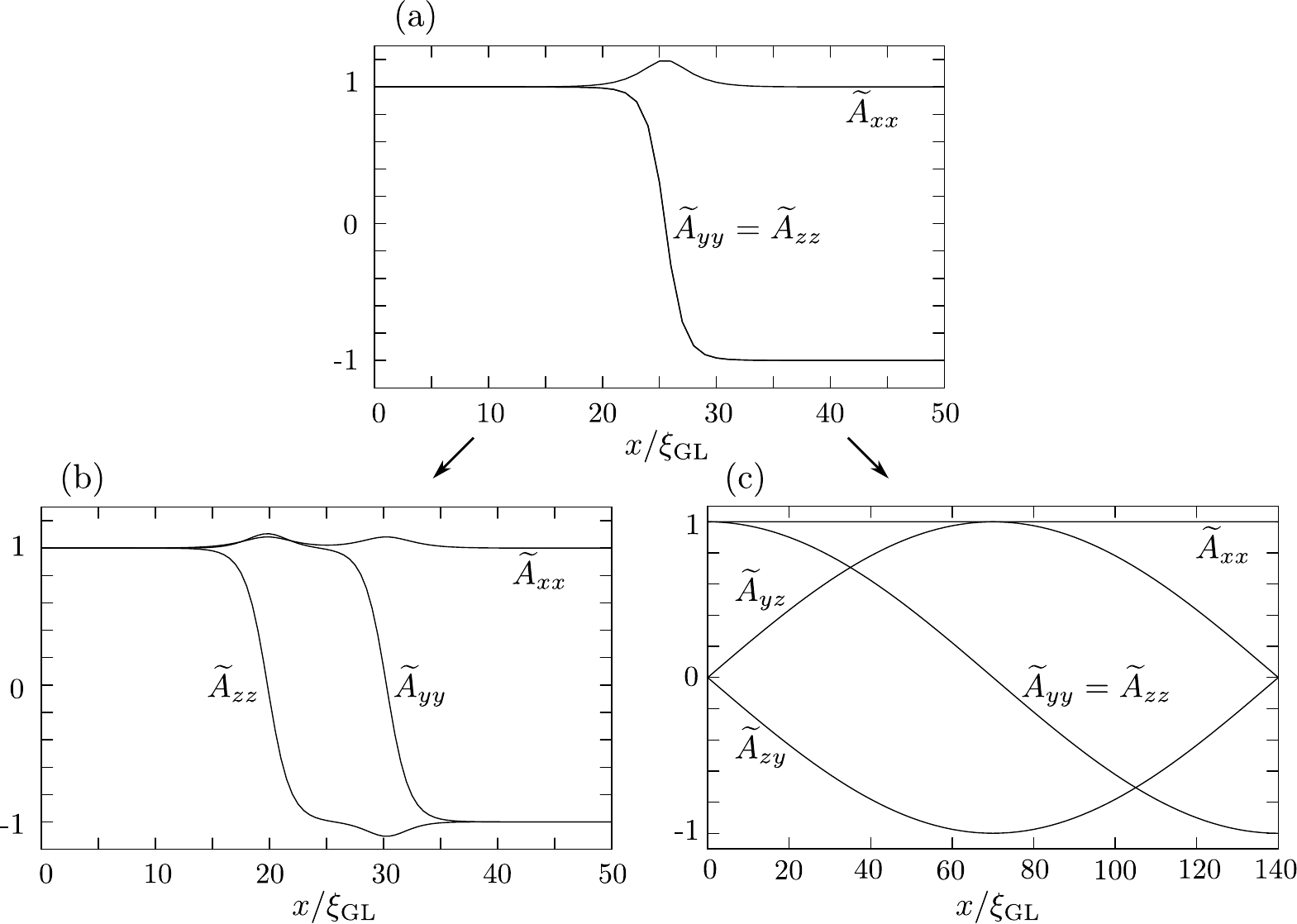}
\caption{\label{BB1m2.fig} Three solutions for B-B-$1\overline{2}$. The single interface (a) decays by minimal perturbation to either a double~interface~(b) or to a texture~(c). The single interface~(a) could be interpreted as two coincident B-B-$10$ interfaces, one shown in Fig.~\ref{BB10.fig}(a) and the other is a rotated version of it. In the double interface (b) they are separated from each other.}
\end{figure*}

The $1\overline{2}$ interface is characterized by a sign change in two components $\widetilde{\matr{A}}_{yy}$ and $\widetilde{\matr{A}}_{zz}$. The first converged solution, a single interface with $\widetilde{\matr{A}}_{yy}\equiv\widetilde{\matr{A}}_{zz}$, is shown in Fig.~\ref{BB1m2.fig}(a). The component $\widetilde{\matr{A}}_{xx}$ is nearly constant but  has enhancement which is approximately twice as large as in  B-B-$10$. A question of independent interfaces emerges: Is B-B-$1\overline{2}$ a unique domain wall or a superposition of two B-B-$10$ domain walls. We find two different modes by which the single interface structure can decay by a low energy perturbation.  With larger activation energies (on the scale of $f_{\rm c}^{\rm B} \xi_{\rm GL}$) also other decay modes are possible, but are omitted here. 

A small perturbation to either $\widetilde{\matr{A}}_{yy}$ or $\widetilde{\matr{A}}_{zz}$ leads to the separation of the two B-B-$10$ domain walls. The minimum amplitude of the perturbation is $C\approx 10^{-5}$. The solution consists of two successive B-B-$10$ domain walls, shown in Fig.~\ref{BB1m2.fig}(b). The double interface should be stable by the stability arguments of the B-B-$10$. With this mode in mind we could interpret B-B-$1\overline{2}$ as a double interface.

The other decay mode leads to the pure texture solution shown in Fig.~\ref{BB1m2.fig}(d) and  analytically written as
\begin{equation}
\widetilde{\matr{A}}_{i j}(x)=\matr{R}_{i j}(\unitvekt{x},-\pi x/L).
\label{BB1m2rot}
\end{equation}
The perturbation needed to break the single interface to this texture solution is extremely small, minimum amplitude $C\gtrsim 10^{-40}$ but the shape must be at least somehow faithful to the final solution. 
The comparison between the minimum amplitudes between the different decay modes reveals that B-B-$1\overline{2}$ is more likely to disintegrate to a texture solution than to exist as a stable double interface structure.
\subsection*{B-B-$\overline{1}0$}

\begin{figure*}
\centering
\includegraphics[width=0.79\linewidth]{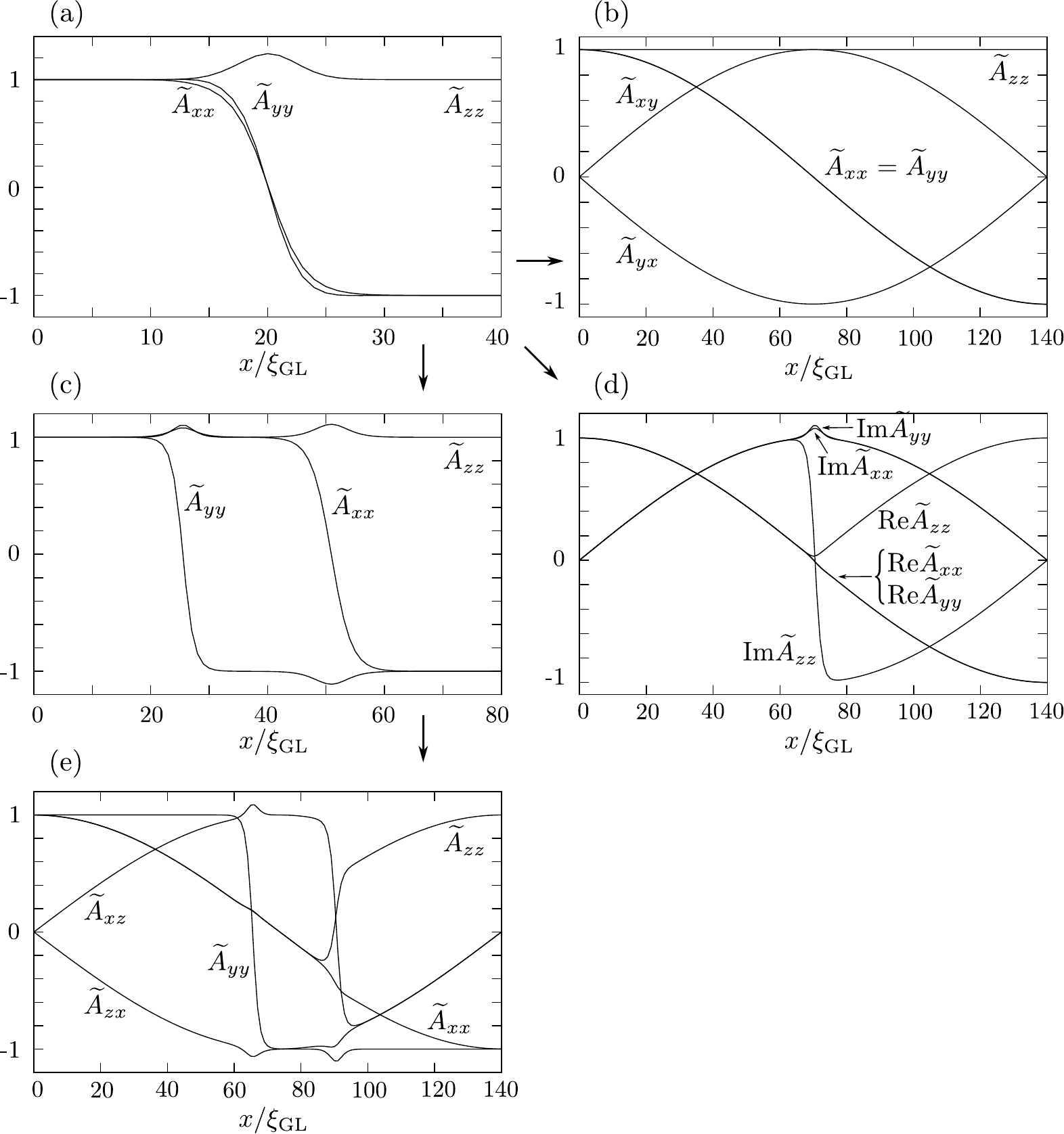}
\caption{\label{BBm10.fig} Solutions for B-B-$\overline{1}0$: The  single interface (a) is a saddle point of energy and decays to a double domain wall (c) of $10$ and $\overline{1}2$. By decay of $\overline{1}2$ this  further deforms to a mixed double interface (e). A perturbation at the single-domain-wall stage can also lead to a mixed solution (d) or to a texture (b). 
}
\end{figure*}

The $\overline{1}0$ interface is characterized by sign changes in the components $\widetilde{\matr{A}}_{xx}$ and $\widetilde{\matr{A}}_{yy}$. The iteration first seems to converge towards the single domain-wall shown in Fig.~\ref{BBm10.fig}(a). One can notice the difference in the slopes of $\widetilde{\matr{A}}_{xx}$ and $\widetilde{\matr{A}}_{yy}$, which can be understood by the anisotropy of the gradient energy discussed above. Also the enhancement of the idle component $\widetilde{\matr{A}}_{zz}$ in the middle is stronger than in the previous analogous structures shown in (a)-panels of the Figures \ref{BB10.fig}-\ref{BB1m2.fig}. A question about the independent interfaces can be posed similarly to the case of B-B-1$\overline{2}$. Now the constituents would be the domain walls $10$ and $\overline{1}2$. The single interface is a saddle point of energy, and we discuss three distinct decay modes below.

Even without any perturbation, the single domain wall disintegrates in further iteration to a double domain-wall consisting of  $10$ and $\overline{1}2$ parts, see Fig.~\ref{BBm10.fig}(c). Due to repulsive interaction, the distance between the parts grows but stops at $\approx 26\: \xi_{\rm GL}$, where their overlap becomes  negligible.  The $10$ domain wall is locally  stable and remains unchanged. The $\overline{1}2$ domain wall deforms further to a mixture of a texture and a $10$ domain wall, as was discussed in connection of  Eq.~\eqref{BB1}. Then there are two built-in $10$ domain walls and a texture in the same B-B-$\overline{1}0$ interface, shown in Fig.~\ref{BBm10.fig}(e). The energy of the solution (Table \ref{TauluKoonti}) is the sum of energies of its constituents $\overline{1}2$ and $10$. The analytic description can be written as
\begin{equation}
\widetilde{\matr{A}}_{ij}(x)=\matr{R}_{i k}(\unitvekt{y},\pi x/L) \widetilde{\matr{A}}_{kl}^{\text{\ref{BB10.fig}(a)}}(x-c_1) \widetilde{\matr{A}^\prime}_{lj}^{\text{\ref{BB10.fig}(a)}}(x-c_2),
\label{BB10.refdoub}
\end{equation}
where $\widetilde{\matr{A}}_{kl}^{\text{\ref{BB10.fig}(a)}}$ is the reduced order parameter of the $10$ domain wall in Fig.\ \ref{BB10.fig}(a), $\widetilde{\matr{A}^\prime}_{lj}^{\text{\ref{BB10.fig}(a)}}$ is the rotated version of this and $c_i$'s are constants. 

The other decay mode of the single domain wall is to a mixed solution of a texture having phase winding  and a $10$ domain wall [Fig.~\ref{BBm10.fig}(d)]. The analytic description is
\begin{equation}
\widetilde{\matr{A}}_{ij }(x)=\erik{e}^{\erik{i} \pi x/L}\widetilde{\matr{A}^\prime}_{ij}^{\text{\ref{BB10.fig}(a)}}\left(x-c\right).
\label{BB4mix}
\end{equation}
The perturbation needed to trigger the deformation toward this solution is mid-strength, amplitude $C \gtrsim 0.1$, and the shape must be designed for the form \eqref{BB4mix}.

The  third decay mode of  the single domain wall  is to a pure texture, shown in Fig.~\ref{BBm10.fig}(b). The perturbation has to  have a shape faithful to the solution and the minimum amplitude $C\approx 10^{-4}$. Analytically this texture is  similar to the pure texture solution of B-B-1$\overline{2}$~\eqref{BB1m2rot}, except that the rotation axis is now $\unitvekt{z}$ instead of $\unitvekt{x}$.

\subsection*{B-B-$\overline{1}\overline{2}$}

\begin{figure*}
\centering
\includegraphics[width=0.8\linewidth]{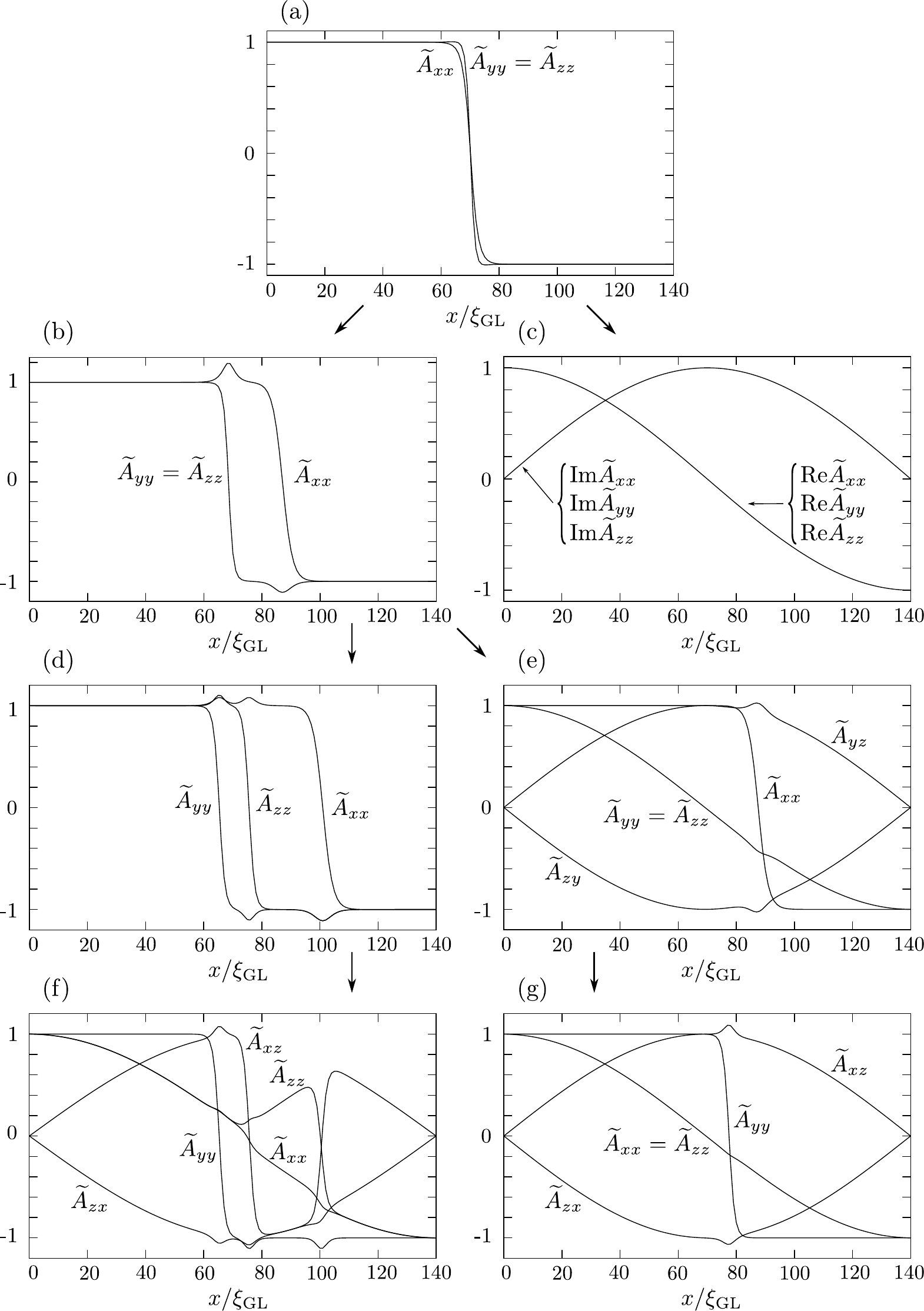}
\caption{\label{BBm1m2.fig} Solutions for the B-B-$\overline{12}$. The single domain wall (a) disintegrates spontaneously to a double one (b), which decays either to a triple domain wall (d) or to a mixed~1 state (e). With minimal perturbation the latter two can turn to  a mixed triple interface (f) and a mixed 2 interface (g), respectively. The single domain wall can also decay directly to a  phase texture (c). The structures (b) and (d)-(g) are combinations of the structures introduced in Figs.~\ref{BB10.fig}-\ref{BBm10.fig}. }
\end{figure*}

The $\overline{1}\overline{2}$ interface is characterized by a sign change of all diagonal components of the order parameter $\widetilde{\matr{A}}_{ij}$. The B-B-$\overline{12}$ could also be called a pure phase wall. First the iteration converges toward a single domain wall, shown in Fig.~\ref{BBm1m2.fig}(a). With further iteration this splits spontaneously to a double domain wall, shown in Fig.~\ref{BBm1m2.fig}(b) and also found in Ref.~\onlinecite{SV}. We find that this structure deforms further with a very small activation energy. 

The double domain wall consists of $1\overline{2}$ and $\overline{1}2$ parts.  As they are $\approx 26 \: \xi_{\rm GL}$ apart, they can  decay independently of each other. The B-B-$1\overline{2}$ can disintegrate to a double domain wall [Fig.~\ref{BB1m2.fig}(b)] producing a triple domain wall [Fig.~\ref{BBm1m2.fig}(d)]. Alternatively it can reduce to a pure texture  [Fig.~\ref{BB1m2.fig}(c)] producing the mixed 1 configuration shown in Fig.~\ref{BBm1m2.fig}(e). Both these are still intermediate states as the $\overline{1}2$ part disintegrates in response to  a minimal perturbation. The triple domain wall deforms to a structure which has three successive $10$ domain walls and a background texture [Fig.~\ref{BBm1m2.fig}(f)]. As a combination of three mutually repulsive domain walls, this structure should be metastable.  Similar decay in the mixed 1 structure produces the mixed state 2 shown in Fig.~\ref{BBm1m2.fig}(g). This is similar to the mixed 1 solution except that the domain wall in the middle is of type $10$ instead of type $\overline{1}2$. The mixed 2 configuration is similar as found in the double-core vortex on the axis passing between the two cores~\cite{EVT}. The mixed~2 is also obtained directly from the single domain wall as a result of a properly designed perturbation with minimum~amplitude~$C\approx 10^{-15}$. 

An alternative decay channel of the single domain wall is to a pure phase texture [Fig.~\ref{BBm1m2.fig}(c)]. This is produced by an imaginary perturbation to diagonal elements of the $\widetilde{\matr{A}}_{ij}$ at minimum amplitude $C\gtrsim 10^{-4}$. The analytic form of the phase texture is 
\begin{equation}
\widetilde{\matr{A}}_{i j}(x)=\erik{e}^{\erik{i}\pi x/L} \matr{\delta}_{i j}.
\label{BB7phase}
\end{equation}

\section{Adiabatic A to B transition}\label{s.adiab}

\begin{figure}
\centering
\includegraphics[width=1.0\linewidth]{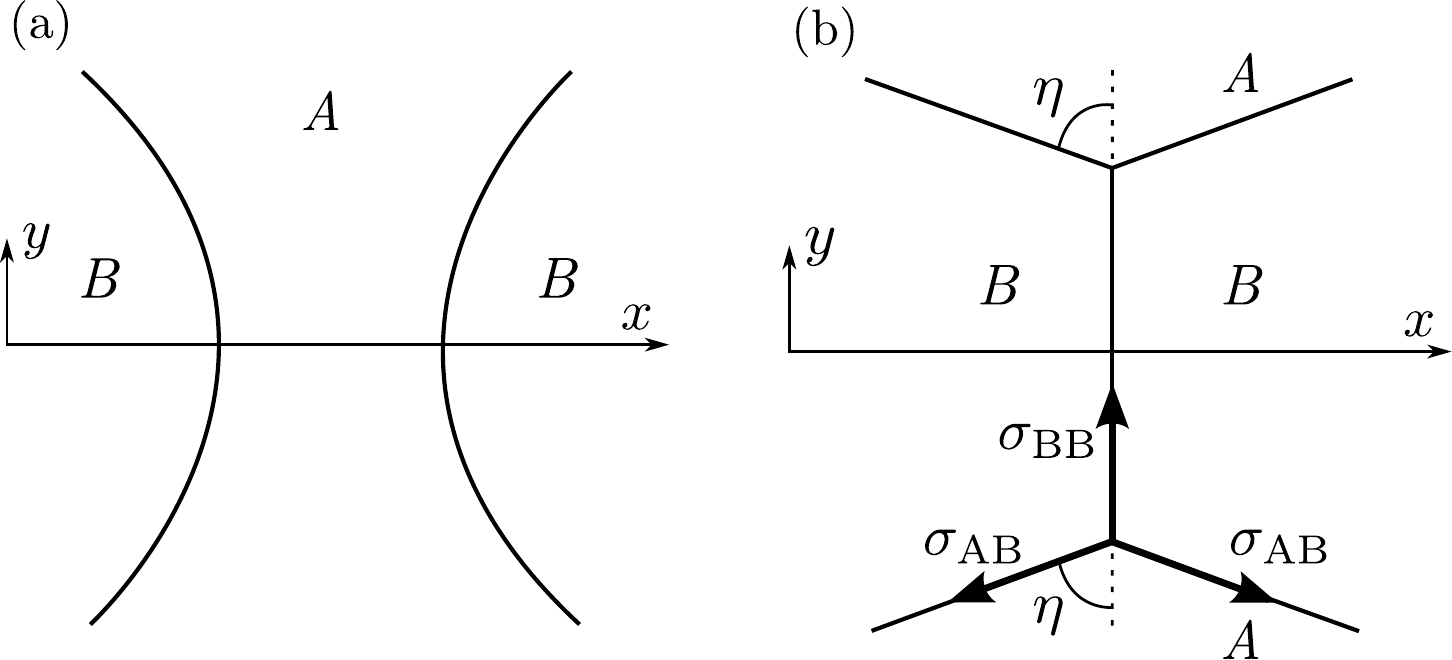}
\caption{\label{BAB2} A $\to$ B phase transition scenario. The panel (a) shows the A phase sandwiched between two  independent B phase patches. In panel (b) the B phase patches have joined forming a B-B domain wall. Also shown are the tension forces and the contact angle at the point where the B-B domain walls terminates to the  A-B interface.}
\end{figure}

In this section we sketch how a hard B-B domain wall could be created in an adiabatic A$\rightarrow$B transition.  For that we  consider a cell of ${}^3$He initially in the A phase. Suppose that upon cooling the B phase has independently nucleated at different locations of the cell, and upon further cooling the patches of B phase expand adiabatically. In this process one finds a shrinking A phase region sandwiched between two independent B phases. The situation is depicted in Fig.~\ref{BAB2}(a).
We now construct the most general order parameter along the $x$ axis, which goes through both A-B interfaces. 

We define the reduced order parameter \eqref{reduc1} using the order parameter on the left B phase, ${\matr{A}}_{ij}^{\rm L}$. As above, this  means that $\widetilde{\matr{A}}_{ij}^{\rm L}$  is equal to the unit matrix. The requirement of a minimum interface energy in the left B-A interface together a specific choice for the direction of the $y$-axis determines that the  reduced order parameter in the A-phase has to be
\begin{equation}
\widetilde{\textsf{A}}^{\rm M}=\widetilde{\Delta}_A\begin{pmatrix} 1 & 0 & \erik{i}\\0&0&0\\0&0&0 \end{pmatrix}
\label{A-phase.middle}
\end{equation}
with $\widetilde{\Delta}_A\sim 1$.
The reason for this is that the anisotropy of the gradient energy \eqref{functional} dictates that the component $\widetilde{\matr{A}}_{xx}(x)$ should change as little as possible and thus $\widetilde{\matr{A}}_{xx}\sim 1$. The structure of the A phase then requires an imaginary $\widetilde{\matr{A}}_{xy}$ or $\widetilde{\matr{A}}_{xz}$ (or combination), and Eq.\ \eqref{A-phase.middle} follows by an appropriate rotation of $y$ and $z$ axes. These boundary conditions on the A-B interface coincide with those in Refs. \onlinecite{EVT2} and \onlinecite{MCC}.
Next we apply the same principles to the A-B interface on the right. Since the axes are now fixed, one should allow an arbitrary rotation around $x$. Moreover, also improper rotations are possible. Instead of explicitly using improper rotations, we represent them with an additional phase factor as in Eq.\ \eqref{order}. Thus we write the two alternative classes of solutions as 
\begin{equation}
\widetilde{\matr{A}}_{ij}^{\rm R}=\begin{cases} \matr{R}_{i j}(\unitvekt{x},\theta) & \text{case 1,}
\\
e^{i\pi}\matr{R}_{ik}(\unitvekt{y},\pi) \matr{R}_{k j}(\unitvekt{x},\theta)&\text{case 2.}
\end{cases}
\label{B-phase.right}
\end{equation}
In both cases $\theta$ can have any value.
\begin{figure*}
\centering
\includegraphics[width=0.8\linewidth]{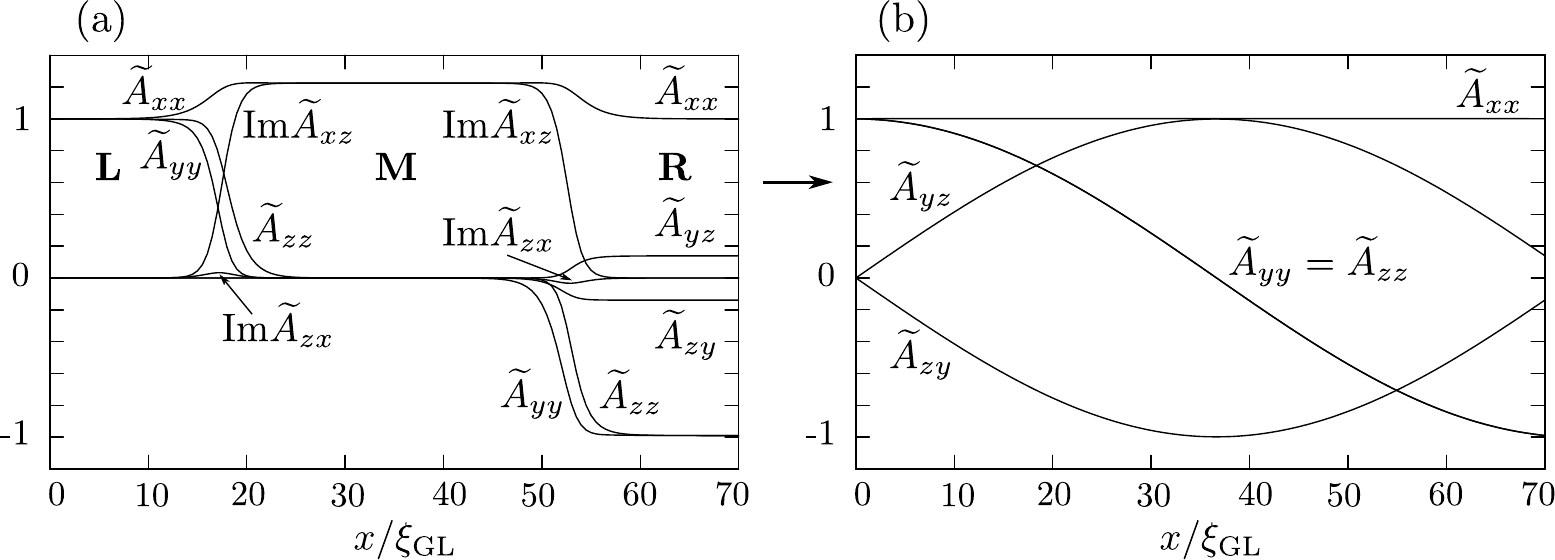}
\caption{\label{BAB::fig:texture} The evolution of B-A-B structure in case 1: (a)  an initial  order parameter configuration in coexistence conditions of A and B phases in case 1 of  \eqref{B-phase.right}  and (b) the final B-phase texture without domain walls.}
\end{figure*}
\begin{figure*}
\centering
\includegraphics[width=0.8\linewidth]{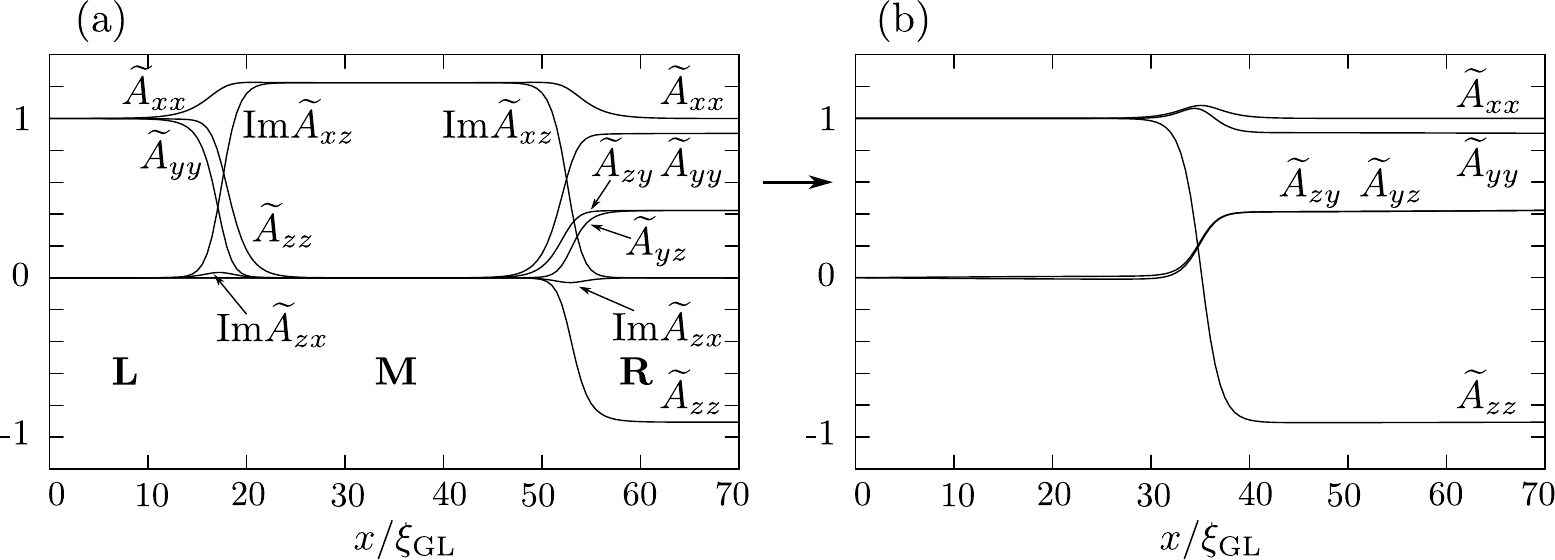}
\caption{\label{BAB::fig:improp:domainwall}The evolution of B-A-B structure in case 2: 
(a) an initial  order parameter configuration in coexistence conditions of A and B phases  in case 2 of  \eqref{B-phase.right}  and  (b) the final B phase with a B-B-$10$ domain wall.}
\end{figure*}

With further cooling the sandwiched A  phase tends to vanish completely. Assuming that this also takes place adiabatically, we can predict the outcome. The result, which is explained in more detail below, is that no  B-B domain wall is expected to form in  case 1 of Eq.\ \eqref{B-phase.right} while a B-B domain wall is always formed in  case 2. For independently nucleated B phases the two cases are equally probable. Thus we expect creation of a B-B domain wall with the probability  of one half. 

We have studied the formation of domain walls using the same 1D numerical relaxation as in the stability studies.   
In case 1 the formation of a B-B domain wall is possible only when the rotation angle $\theta$ is in the region $\pi \pm \epsilon$, where $\epsilon/\pi \ll 1$. If the order parameter does not meet this constraint, a B-phase texture is formed instead of any domain wall, as shown in Fig.~\ref{BAB::fig:texture}. Without a method to control the angle $\theta$, the probability of a B-B domain wall formation in case 1 is nearly zero. In case 2  the evolution always ends  with a B-B domain wall. The evolution in the latter case is demonstrated in the Fig. \ref{BAB::fig:improp:domainwall}. 

The domain wall, when created, is in contact with the A-B interface. At the contact line the tensions of the interfaces must balance. Defining the a contact angle $\eta$ as indicated in Fig.\ \ref{BAB2}(b), one gets $\cos \eta=\sigma_{\rm BB}/2\sigma_{\rm AB}$, where  
$\sigma_{\rm BB}$ and $\sigma_{\rm AB}$ are the tensions of the B-B domain wall and the A-B interface, respectively. We have calculated the  tensions \eqref{sigma} numerically using the GL theory, which then allows to obtain $\eta$. When  magnetic field is applied, the A-B interface can be stabilized in the validity range of the GL theory at all pressures below the polycritical pressure. Therefore, we can determine the contact angle as a function of pressure, $\eta(p)$. The results are given in  Table \ref{TauluWettingAngle}.

In the weak-coupling limit all the intermediate states in the A-B interface and the B-B-10 domain wall become degenerate \cite{volovik90,EVT2}. Thus both turn from well defined domain walls to textures, and their tensions vanish. Based on our calculation  $\sigma_{\rm BB}$ vanishes more rapidly than  $\sigma_{\rm AB}$, and therefore $\eta$ approaches $90^\circ$ in the weak-coupling limit (which corresponds to zero pressure on our scale).
\begin{table}[t]
\begin{center}
\begin{tabular}{lcccccccc}
\toprule
Pressure (bar) &$0.5$&$1$&$3$&$6$&$12$&$16$&$24$&PCP\\
\midrule
Contact angle $\eta$ &$89^\circ$&$80^\circ$&$77^\circ$&$74^\circ$&$70^\circ$&$68^\circ$&$65^\circ$&$63^\circ$\\
\bottomrule
\end{tabular}
\end{center}
\caption{\label{TauluWettingAngle} The contact angle $\eta$ between A-B interface and B-B-10 domain wall calculated using Ginzburg-Landau theory. At pressures lower than the polycritical one the coexistence of A and B phases is achieved by applying a magnetic field.  For details of the theoretical pressure scale see Section \ref{s.GL}.
}

\end{table}

\section{Dipole-dipole interaction energy in the domain wall}\label{s.dipole}

Until now we have studied the interfaces on the condensation-energy scale. This sets conditions on the  the reduced order parameter $\widetilde{\matr{A}}_{ij}(\bm r)$, but it leaves the full order parameter \eqref{reduc1} undetermined by a rotation matrix ${\matr{R}}_{ij}(\unitvekt{n}, \theta)$.  In order to constrain these soft degrees of freedom, we have to look at weaker contributions to the energy. In particular, we consider the dipole-dipole energy
\begin{equation}
F_{\rm D}=g_{\rm D}\int \D^3r(\matr{A}^*_{ii}\matr{A}_{jj}+\matr{A}^*_{ij}\matr{A}_{ji}-\frac{2}{3}\matr{A}^*_{ij}\matr{A}_{ij}).
\label{functionaldipole}
\end{equation}
Because $g_{\rm D}$ is small compared to $\alpha$ (except at temperatures very close to $T_c$), the dipole-dipole energy in a domain wall itself is negligible, but it is important in the bulk on both sides of the the domain wall. On the left hand side the order parameter $\matr{A}_{\mu i}=\Delta \erik{e}^{\erik{i}\phi}\matr{R}_{\mu i}(\unitvekt{n}, \theta)$. The minimization of the dipole energy \eqref{functionaldipole} for this order parameter  leads to locking of the rotation angle $\theta$ to $\theta_{\rm L}=\arccos(-1/4)\approx 104^\circ$. This sets no constraints  on the  rotation axis $\unitvekt{n}$. 
On the right hand side of the domain wall, the dipole energy  should be minimized by the order parameter  $\matr{A}_{\mu i}=\Delta \erik{e}^{\erik{i}\phi}\matr{R}_{\mu j}(\unitvekt{n}, \theta_L)\matr{\widetilde{A}}^{\rm R}_{j i}$.  This leads to additional conditions if $\matr{\widetilde{A}}^{\rm R}_{j i}$ (see Table \ref{TauluKoonti}) is not proportional to the unit matrix. 

We concentrate on the most stable domain wall B-B-10 represented in Fig.~\ref{BB10.fig}(a).  
The minimization gives on the rotation axis $\unitvekt{n}=(\hat n_x,\hat n_y,\hat n_z)$ the constraint
\begin{eqnarray}
\hat n_y=\pm\sqrt{\frac{3}{5}}.
\label{dipoleN10}
\end{eqnarray}
or $\hat n_x^2+\hat n_z^2=2/5$.
Considering the minimization of  the dipole energy on the $(\unitvekt{n},\theta)$-ball\cite{VM}, the first minimization reduces the full $\pi$ radius ball to a $\theta_{\rm L}$ radius sphere and the second minimization reduces further the sphere to two $\sqrt{2/5}$ radius circles perpendicular to the $y$-axis, midpoints lying on the $y$-axis at the distance $\sqrt{3/5}$ from origin.  As the $y$ axis is in the plane of the domain wall, the component of $\hat{\bm n}$ on the wall normal is always less or equal to $\sqrt{2/5}$.

The constraint for $\unitvekt{n}$ is very strong compared to other orienting effects that act  on $\unitvekt{n}$ \emph{inside} the superfluid.  Similar $\unitvekt{n}$-textures are previously known to occur only near boundaries, such as the container wall or the free liquid surface \cite{SBE}. These $\unitvekt{n}$-textures should be observable in NMR-experiments if the effect from nearby walls could be suppressed and the relative volume of the metastable B-B domain wall is sufficiently large. 

\section{Domain wall in contact with surfaces}\label{s.surf}

\begin{figure}[tb] 
   \centering
   \includegraphics[width=0.8\linewidth]{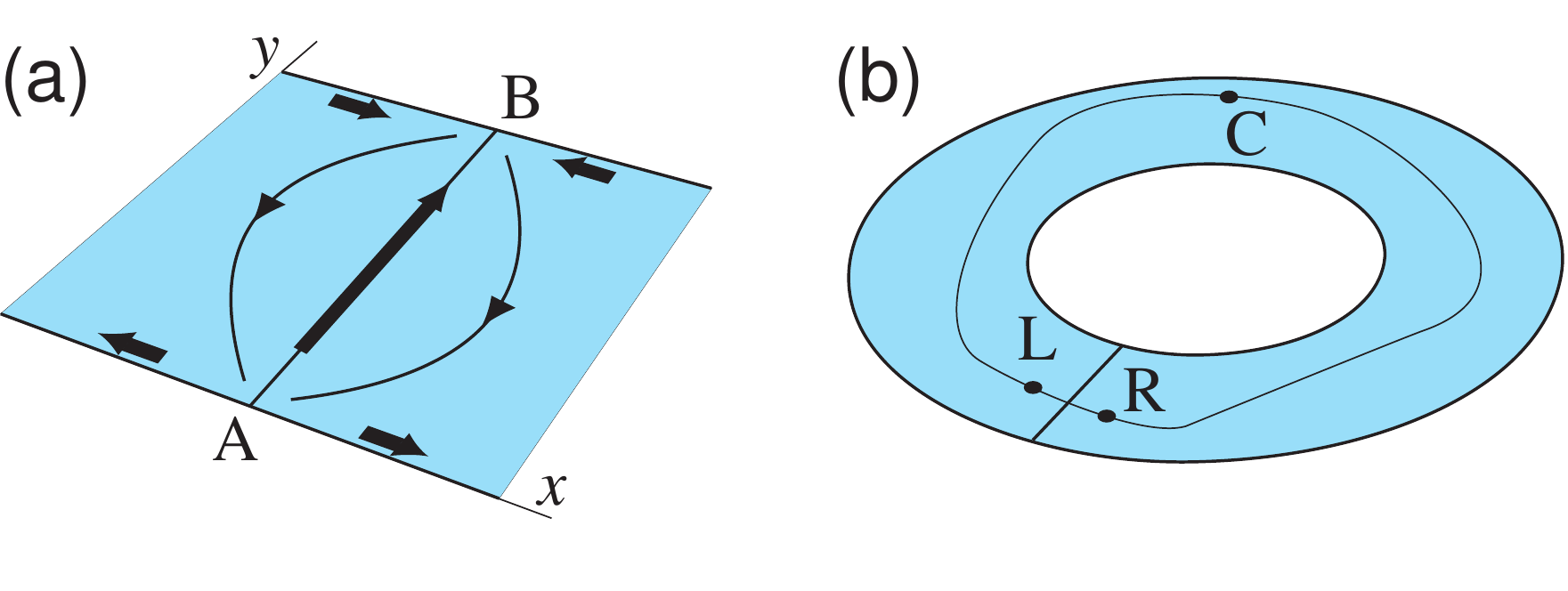} 
   \caption{ (a) A B-B-10 interface having contact lines with two sample walls. The figure shows a 2D projection, where the contact lines appear as points A and B. The arrows indicate spin flow $(\widetilde J_{zx}, \widetilde J_{zy})$ in the $x-y$ plane. The thick arrows show spin currents in the domain wall and on the surfaces. The curved arrows depict spin current in the bulk that is necessary to satisfy spin current conservation.  (b) The domain wall in a container topologically equivalent to a torus and a topologically nontrivial closed path passing through points L, R and C.}
   \label{f.schemas}
\end{figure}

Consider superfluid $^3$He limited by two planar walls at $y=0$ and $y=L_y$. In such a geometry a B-B-10 interface can be locally stable stretching from one wall to the other, and has  contact lines with both walls, see Fig.\ \ref{f.schemas}(a). We have studied this situation by solving numerically the GL equations in two dimensions. At the walls we use the boundary condition that all components of the order parameter vanish. This  approximately mimics real surfaces where quasiparticles are scattered diffusely. We have studied the range
of $L_y/\xi_{\rm GL}$ from 20 to 130. We neglect the effect of dipole-dipole interaction, which should be a good approximation at length scales smaller than the dipole length $\xi_D\sim10\ \mu$m.

\begin{figure*}
\centering
\includegraphics[width=0.475\linewidth]{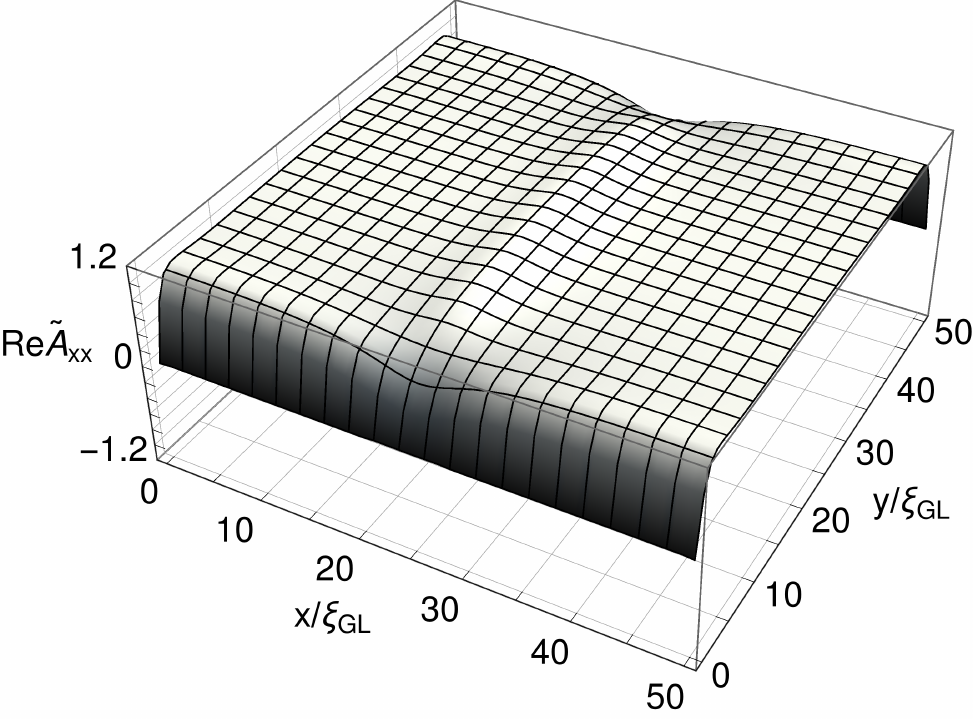}
\hspace*{0.02\linewidth}
\includegraphics[width=0.475\linewidth]{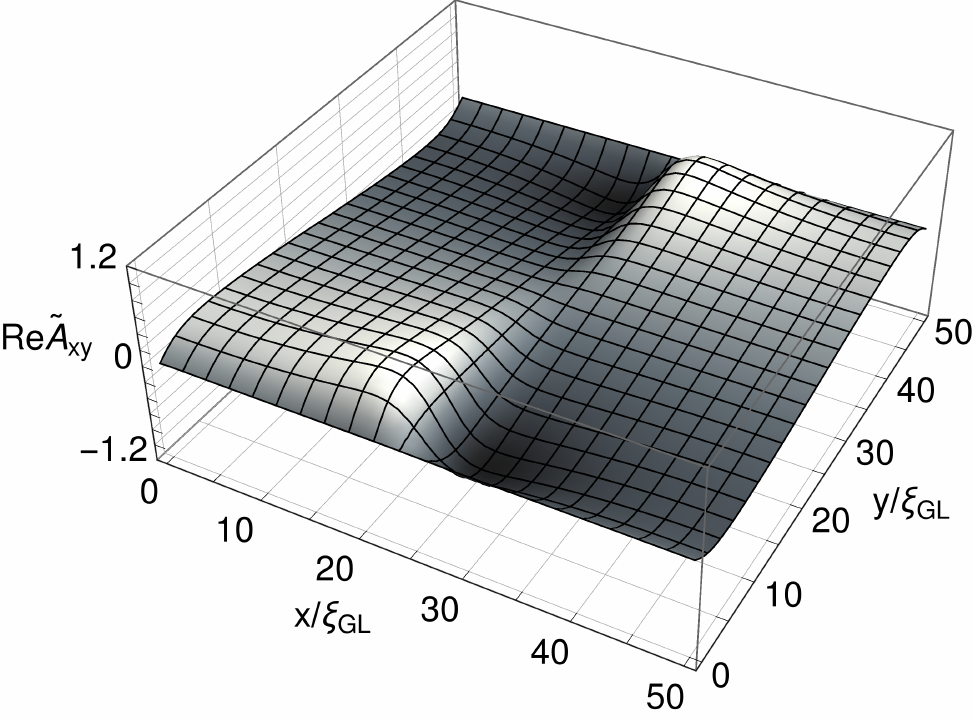}
\includegraphics[width=0.475\linewidth]{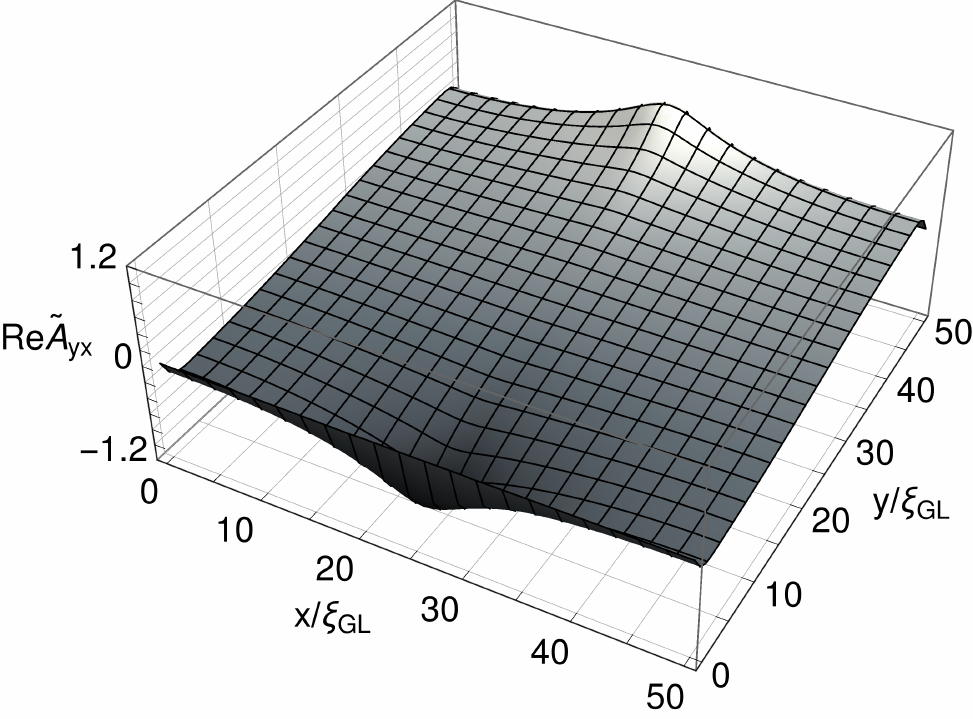}
\hspace*{0.02\linewidth}
\includegraphics[width=0.475\linewidth]{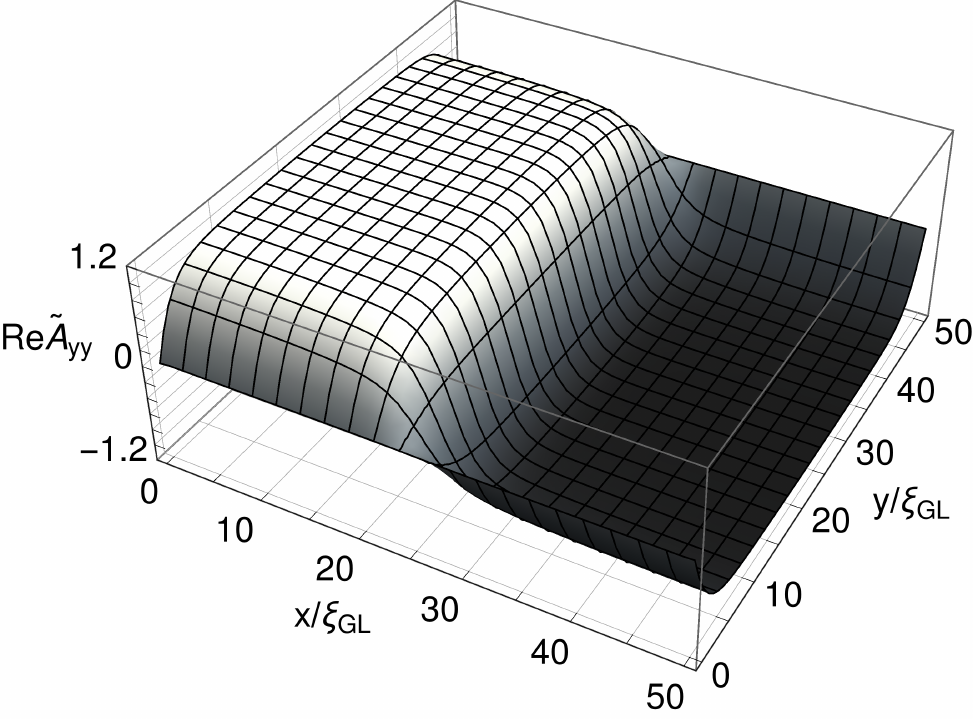}
\includegraphics[width=0.475\linewidth]{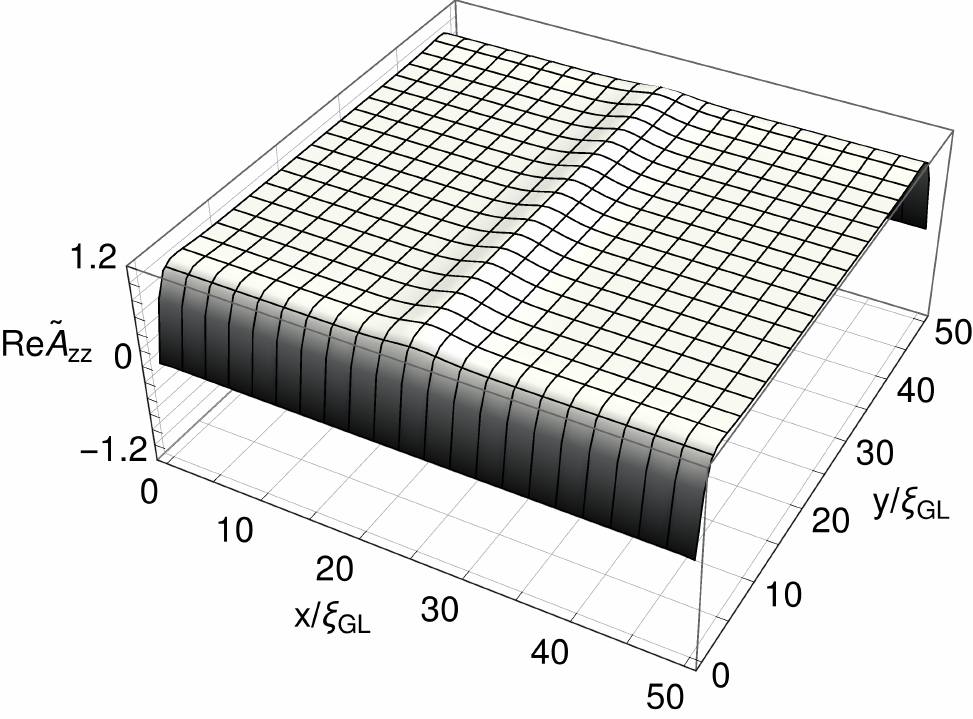}  
\hspace*{0.02\linewidth}  
\includegraphics[width=0.475\linewidth]{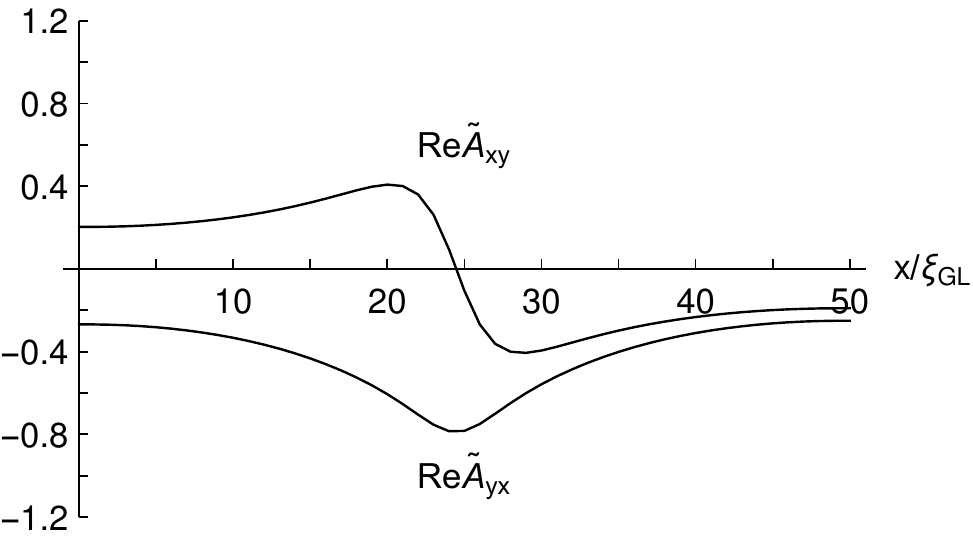}  
\caption{The order parameter of minimum energy  B-B-10 interface
stretching between two solid walls. The sample walls are at $y=0$ and
$y=L_y=50\xi_{\rm GL}$. Only nonvanishing components of the order
parameter are shown. The panel at bottom right shows $\widetilde A_{xy}$
and $\widetilde A_{yz}$ plotted along the $x$ axis at $y=4\xi_{\rm
GL}$.}
\label{f.Kanavajp}  
\end{figure*}

One question of interest is to determine the orientation of the interface gap node with respect to the walls. We find the lowest energy when the gap-node direction is perpendicular to the wall. The order parameter corresponding to this solution is shown in Fig.\ \ref{f.Kanavajp}.

A striking feature of the order parameter is the appearance of 5 real non-zero components. For both the state on a wall as well as for the B-B-10 interface, only the diagonal components are present. The appearance of the extra components $\widetilde A_{xy}$ and $\widetilde A_{yz}$ can be understood by studying spin currents. The spin current in the interface was studied above in connection of Eq.\ (\ref{e.spiczy}). The surface state also has spin current \cite{ZKT}. It is given by the tensor $\textsf{J}_{\rm surf}=c_{\rm surf}\textsf{R}(\hat{\bm u}\,\hat{\bm t}-\hat{\bm t}\,\hat{\bm u})$. Here $\hat{\bm u}=\hat{\bm s}\times \hat{\bm t}$, $\hat{\bm s}$ is the surface normal pointing to the superfluid, $\hat{\bm t}$ an arbitrary unit vector perpendicular to $\hat{\bm s}$, and $c_{\rm surf}=0.238\times 4K\Delta2/\hbar$.  Applying these to the case of Fig.\ \ref{f.Kanavajp} gives that $\widetilde J_{z i}$ in the interface as well as in the surface states on  both sides of a contact line are all towards one contact line and away from the other [Fig.\ \ref{f.schemas}(a)].
Since the spin current has to be conserved (assuming negligible the dipole-dipole interaction), there has to be compensating spin currents around the interface. 
The components $\widetilde A_{xy}$ and $\widetilde A_{yz}$ appear just to generate this compensating spin current so that the  conservation law $\partial_x\widetilde J_{z x} +\partial_y\widetilde J_{z y} =0$ holds.

The structure in Fig.\ \ref{f.Kanavajp} is closely related to the striped phase discovered by Vorontsov and Sauls \cite{Vorontsov07}. They studied the parallel plate geometry of thickness on the order of the coherence length using specular boundary condition. They found that a periodic B-B-10 interface structure can be the ground state of the system. Similarly as above, the appearance of $a_{zx}$ component in Ref.\ \onlinecite{Vorontsov07} is caused by  spin current conservation.

The interface structure in Fig.\ \ref{f.Kanavajp} could be compared to the one where the gap node is in the $z$ direction. This structure has no bulk spin currents since in the interface only $\widetilde J_{y z}$ is nonzero and $\widetilde J_{z x}$ in the surface state is in the same direction on both sides of a contact line. The reason for the higher energy of this state is the larger energy associated with the contact line: the gradient energy associated with $\partial_y\widetilde A_{yy}$ in this state is by factor 3 more costly than the one with $\partial_y\widetilde A_{zz}$ in the structure of Fig.\ \ref{f.Kanavajp}.

\section{Quantization of circulation}\label{s.circ}

Evidence of $\pi$-shift of circulation in superfluid $^3$He-B was found in the experiment of Ref.\ \onlinecite{MAV}. The purpose of this section is to show that a B-B-10 domain wall in the flow path can give rise to this observation. Moreover, as  B-B-10 is the only locally stable structure that can do this, we reach a unique identification of B-B-10 in this experiment.
The $\pi$-shift of circulation appears for interfaces having $\phi=\pi$, see Table \ref{TauluKoonti}. For completeness, we go through the argument in more detail below. 

Let us consider a closed path in superfluid $^3$He-B. Assuming the order parameter has the bulk  form (\ref{order}) everywhere on the path, the phase change $\Delta\phi=\oint\bm\nabla\phi\cdot d\bm r$ on traversing the path has to be $2\pi n$ with integer $n$. This is commonly expressed by saying that the circulation is quantized to integer values. 

Let us study B phase in a container that is topologically equivalent to a torus, see Fig.\ \ref{f.schemas}(b).  Consider a topologically nontrivial path that circles the torus once. The minimal object that can lead to deviation from the integer quantization rule is  a hard domain wall so that  the path passes through it once. On the path outside of the domain wall (from L to R via C), the order parameter on the left hand side of the domain wall has to change smoothly to the one on the right hand side  keeping the bulk  form (\ref{order}). This leads to the condition
\begin{eqnarray}
e^{i\Delta\phi}\widetilde{\textsf{A}}^R=\textsf{R},
\label{e.cicorbb10}\end{eqnarray}
where  $\textsf{R}$ is a proper rotation matrix. For B-B-10 we have $\widetilde{\textsf{A}}^R={\rm diag}(1,-1,1)$ (Table \ref{TauluKoonti}). Since $\textsf{R}$ is real and $\det \textsf{R}=1$, it follows from (\ref{e.cicorbb10}) that $\Delta\phi=\pi+2\pi n$ with integer $n$. Thus for the locally stable interface B-B-10 there is a $\pi$ shift in the quantization, which allows half-quantum circulation. The same quantization rule applies also to interfaces of types $\overline{1}2$ and $\overline{1}\overline{2}$, but these are not locally stable structures.

\section{Bound quasiparticle states}\label{s.excitation}

The quasiparticle excitation spectrum can be studied either by solving Bogoliubov-de Gennes equations or by Green's function methods.  Calculations close to the present case are reported in Refs.\ \onlinecite{Kulik69,Nakahara86,Nagai08,Tsutsumi12,Mizushima12,Silaev12,Wu13}. Similar to earlier work, we find bound states below the gap energy. These Andreev bound states can be interpreted to consist of a superposition of a particle-like and a hole-like excitations that have nearly the same momentum but travel in opposite directions. The bound state has no net mass because the hole and particle masses are opposite. The bound states transport mass, but in equilibrium the net current vanishes in a B-B-10 interface. In spin-triplet superfluids the bound states have no spin as the spins of the particle and hole parts are opposite. The bound states transport spin in spin-triplet superfluids. 

A precise calculation of the bound states requires numerical methods. Instead, we can get a qualitative picture using the following model for  B-B-10. We assume that $\widetilde A_{yy}$ has a step-like change from  $1$ to $-1$ at the domain wall and other components stay constants, $\widetilde A_{xx}=\widetilde A_{zz}=1$. 
We find quasiparticle states with energy   
\begin{eqnarray}
E=\mp\Delta\sqrt{\hat p_x^2+\hat p_z^2}\mathop{\rm Sgn}(\hat p_y).
\end{eqnarray}
Here $\hat {\bm p}$ is the direction of the momentum of the excitation. The upper sign is for excitations that transport spin up particles and the lower sign for spin down transport. The energy vanishes in the gap-node direction $\hat{\bm p}\parallel\hat{\bm y}$. Excitations with negative energy can be interpreted to be filled in the ground state. In B-B-10 these give rise to the spin current discussed in Secs.\ \ref{s.BB10} and \ref{s.surf}.

\section{Conclusion}

We have made stability analysis of different candidate structures of hard B-B interfaces. It is found that only one of these, B-B-10, is a locally stable structure. We have studied the properties of this interface and its nucleation in an A$\rightarrow$B transition.  The observation of half-quantum circulation in the experiment of Ref.\ \onlinecite{MAV} can be interpreted as the presence of B-B-10. Evidence of the interface is also presented in the experiment of Ref.\ \onlinecite{WBG}. We suggest that B-B-10 could also be present in the experiment of  Ref.\ \onlinecite{Brad}: it could be stabilized in the high-field measuring region due to  a positive magnetic susceptibility eigenvalue \eqref{BB10susc}, and be responsible for the reduction of quasiparticle transmission. What still remains for future experiments is to localize the defect. This  probably requires measurement of the texture, possibly by NMR, in a properly designed geometry that can trap the domain wall.

\begin{acknowledgments}
We thank Yury Bunkov and Yury Mukharsky for discussions and Joonas Keski-Rahkonen for corrections.
This work was financially supported by the  Academy of Finland.
\end{acknowledgments}



\begin{thebibliography}{99}

\bibitem{Maki} K. Maki and P. Kumar, \textit{Phys. Rev. B} {\bf 14}, 118 (1976), \textit{Phys. Rev. Lett.} {\bf 38}, 557 (1977).

\bibitem{Hanninen03} R. H\"anninen and E. V. Thuneberg, \textit{Phys. Rev. B} {\bf  68}, 094504 (2003).

\bibitem{smvp} J. S. Korhonen, Y. Kondo, M. Krusius, E. V. Thuneberg and G. E. Volovik, \textit{Phys. Rev. B} {\bf 47}, 8868 (1993).

\bibitem{Levitin13} L. V. Levitin, R. G. Bennett, E. V. Surovtsev,  J. M. Parpia, B. Cowan, A. J. Casey and J. Saunders, Phys. Rev. Lett. {\bf 111},  235304 (2013).

\bibitem{abinterface} A. J. Leggett and S. K. Yip, in \textit{Helium Three}, edited by W. P. Halperin and L. P. Pitaevski (Elsevier, Amsterdam, 1990), p. 523.

\bibitem{EVT2} E. V. Thuneberg, \textit{Phys. Rev. B} \textbf{44}, 9685 (1991).

\bibitem{Ohmi82} T. Ohmi, M.  Nakahara, T. Tsuneto, and T. Fujita, Prog. Theor. Phys.   {\bf 68},  1433 (1982).

\bibitem{SV} M. M. Salomaa and G. E. Volovik, \textit{Phys. Rev. B} \textbf{37}, 9298 (1988).

\bibitem{MAV} Yu. Mukharsky, O. Avenel and E. Varoquaux, \textit{Phys. Rev. Lett.} \textbf{92}, 210402 (2004).

\bibitem{WBG} C. B. Winkelmann, J. Elbs, Yu. M. Bunkov and H. Godfrin, \textit{Phys. Rev. Lett.} \textbf{96}, 205301 (2006).

\bibitem{EVT} E. V. Thuneberg, \textit{Phys. Rev. B} \textbf{36}, 3583 (1987).

\bibitem{VT02} J. K. Viljas and E. V. Thuneberg, \textit{Phys. Rev. B} {\bf 65}, 064530 (2002).

\bibitem{Vorontsov07} A. B. Vorontsov and J. A. Sauls, Phys. Rev. Lett. {\bf 98}, 045301 (2007).

\bibitem{Leggett} A. J. Leggett, \textit{Rev. Mod. Phys.} {\bf 47}, 331 (1975).

\bibitem{book} D. Vollhardt and P. W\"olfle, \textit{The Superfluid Phases of Helium 3}, (Taylor\&Francis, London, 1990).

\bibitem{VM} V. P. Mineyev and G. E. Volovik,  \textit{Phys. Rev. B} \textbf{18}, 3197 (1978).

\bibitem{SS} J. A. Sauls and J. W. Serene, \textit{Phys. Rev. B} {\bf 24}, 183 (1981).

\bibitem{VorSau} A. Vorontsov and J. A. Sauls, \textit{J. Low Temp. Phys.}, \textbf{138}, 283 (2005).

\bibitem{AGR} V. Ambegaokar, P. G. de Gennes and D. Rainer, \textit{Phys. Rev. A} {\bf 9}, 2676 (1974), {\bf 12}, 345 (1975).

\bibitem{BJ} L. J. Buchholtz and G. Zwicknagl, \textit{Phys. Rev. B} {\bf 23}, 5788 (1981).

\bibitem{MCC} M. C. Cross, \textit{Quantum Fluids and Solids}, edited by S. B. Trickey, E. D. Adams and J. W. Dufty (Plenum, New York, 1977), p. 183.

\bibitem{volovik90} G.E. Volovik, Pis'ma Zh. Eksp. Teor. Fiz. {\bf 52}, 972 (1990) [JETP Lett. {\bf 52}, 358 (1990)].

\bibitem{SBE} H. Smith, W.F. Brinkman, and S. Engelsberg, Phys. Rev. B {\bf 15}, 199 (1977).

\bibitem{ZKT} W. Zhang, J. Kurkij\"arvi, and E.V. Thuneberg, Phys. Rev. B{\bf 36}, 1987 (1987).

\bibitem{Kulik69} I. O. Kulik,  {\em Zh. Eksp. Teor. Fiz.} {\bf 57}, 1745 (1969)  [{\em Sov. Phys. JETP} {\bf 30}, 944 (1970)].

\bibitem{Nakahara86} M. Nakahara, J. Phys. C {\bf 19}, L195 (1986).

\bibitem{Nagai08} K. Nagai, Y. Nagato, M. Yamamoto, and S. Higashitani, J. Phys. Soc. Jpn. {\bf 77}, 111003 (2008).

\bibitem{Tsutsumi12} Y. Tsutsumi and K. Machida, J. Phys. Soc. Jpn. {\bf 81} 074607 (2012).

\bibitem{Mizushima12} T. Mizushima, Phys. Rev. B {\bf 86}, 094518 (2012).

\bibitem{Silaev12} M. A. Silaev and G. E. Volovik, Phys. Rev. B {\bf 86}, 214511 (2012).

\bibitem{Wu13} H. Wu and J. A. Sauls, Phys. Rev. B {\bf 88}, 184506 (2013).

\bibitem{Brad} D. I. Bradley, S. N. Fisher, A. M. Gu\'enault, R. P. Haley, J. Kopu, H. Martin, G. R. Pickett, J. E. Roberts and V. Tsepelin, \textit{Nature Physics} \textbf{4}, 46 (2008).

\end{thebibliography}
\end{document}